\documentclass[a4paper,fleqn,usenatbib]{mnras}

\usepackage{newtxtext,newtxmath}
\usepackage[T1]{fontenc}
\usepackage{ae,aecompl}

\usepackage{scrextend}
\usepackage[hyphenbreaks]{breakurl}
\usepackage{threeparttable}

\usepackage{graphicx}	% Including figure files
\usepackage{amsmath}	% Advanced maths commands
\usepackage{amssymb}	% Extra maths symbols

\title[High-precision eclipse photometry of WASP-103\,b]{High-precision multi-wavelength eclipse photometry of the ultra-hot gas giant exoplanet WASP-103\,b}

\author[L. Delrez et al.]{L.~Delrez,$^{1,2}$\thanks{E-mail: \url{lcd44@cam.ac.uk}}
N.~Madhusudhan,$^{3}$
M.~Lendl,$^{4,5,6}$
M.~Gillon,$^{2}$
D. R.~Anderson,$^{7}$
\newauthor
M.~Neveu-VanMalle,$^{5}$
F.~Bouchy,$^{5}$
A.~Burdanov,$^{2}$
A.~Collier-Cameron,$^{8}$
\newauthor
B.-O.~Demory,$^{9,1}$
C.~Hellier,$^{7}$
E.~Jehin,$^{2}$
P.~Magain,$^{2}$
P. F. L.~Maxted,$^{7}$
\newauthor
D.~Queloz,$^{1,5}$
B.~Smalley,$^{7}$ 
and A. H. M. J.~Triaud$^{3}$
\vspace{0.1cm}
\\
$^{1}$ Astrophysics Group, Cavendish Laboratory, J.J. Thomson Avenue, Cambridge CB3 0HE, UK\\
$^{2}$ Space sciences, Technologies and Astrophysics Research (STAR) Institute, Universit\'{e} de Li\`{e}ge, all\'{e}e du 6 Ao\^{u}t 17, 4000 Li\`{e}ge, Belgium\\
$^{3}$ Institute of Astronomy, University of Cambridge, Cambridge CB3 0HA, UK\\
$^{4}$ Space Research Institute, Austrian Academy of Sciences, Schmiedlstr. 6, 8042 Graz, Austria\\ 
$^{5}$ Observatoire de Gen\`eve, Universit\'{e} de Gen\`{e}ve, 51 Chemin des Maillettes, 1290 Sauverny, Switzerland\\
$^{6}$ Max Planck Institute for Astronomy, K\"onigstuhl 17, 69117 Heidelberg, Germany\\
$^{7}$ Astrophysics Group, Keele University, Staffordshire, ST5 5BG, UK\\
$^{8}$ SUPA, School of Physics and Astronomy, University of St. Andrews, North Haugh, Fife, KY16 9SS, UK\\
$^{9}$ University of Bern, Centre for Space and Habitability, Sidlerstrasse 5, CH-3012, Bern, Switzerland
}

\date{Accepted XXX. Received YYY; in original form ZZZ}

\pubyear{2017}

\begin{document}
\label{firstpage}
\pagerange{\pageref{firstpage}--\pageref{lastpage}}
\maketitle

\begin{abstract}
We present sixteen occultation and three transit light curves for the ultra-short period hot Jupiter WASP-103b, in addition to five new radial velocity measurements. We combine these observations with archival data and perform a global analysis of the resulting extensive dataset, accounting for the contamination from a nearby star. We detect the thermal emission of the planet in both the $z'$ and $K_{\mathrm{S}}$-bands, the measured occultation depths being 699$\pm$110 ppm (6.4-$\sigma$) and $3567_{-350}^{+400}$ ppm (10.2-$\sigma$), respectively. We use these two measurements together with recently published HST/WFC3 data to derive joint constraints on the properties of WASP-103b's dayside atmosphere. On one hand, we find that the \hbox{$z'$-band} and WFC3 data are best fit by an isothermal atmosphere at 2900 K or an atmosphere with a low H$_2$O abundance. On the other hand, we find an unexpected excess in the $K_{\mathrm{S}}$-band measured flux compared to these models, which requires confirmation with additional observations before any interpretation can be given. From our global data analysis, we also derive a broad-band optical transmission spectrum that shows a minimum around 700 nm and increasing values towards both shorter and longer wavelengths. This is in agreement with a previous study based on a large fraction of the archival transit light curves used in our analysis. The unusual profile of this transmission spectrum is poorly matched by theoretical spectra and is not confirmed by more recent observations at higher spectral resolution. Additional data, both in emission and transmission, are required to better constrain the atmospheric properties of WASP-103b.

\end{abstract}

\begin{keywords}
planetary systems --
planets and satellites: atmospheres --
stars: individual: WASP-103 --
techniques: photometric --
techniques: radial velocities --

\end{keywords}

%%%%%%%%%%%%%%%%%%%%%%%%%%%%%%%%%%%%%%%%%%%%%%%%%%

%%%%%%%%%%%%%%%%% BODY OF PAPER %%%%%%%%%%%%%%%%%%

\section{Introduction}

\defcitealias{gillon14}{G14}
\defcitealias{southworth15}{S15}
\defcitealias{southworth16}{S16}
\defcitealias{wollert2015}{W15}
\defcitealias{ngo16}{N16}
\defcitealias{cartier17}{C17}

Transiting extrasolar planets are key objects for the understanding of worlds beyond our Solar System, as they provide a wealth of information about their systems. Thanks to their special orbital configuration, we can not only measure their radius, true mass (by either radial velocity measurements or via transit timing variations in multi-planet systems), and orbital parameters, but we can also study their atmosphere and thus gain a complete picture of their chemical and physical properties (e.g. \citealt{demingseager}, \citealt{winn}, \citealt{burrowsreview}, \citealt{madhureviewbis}). These atmospheric studies are conducted using mainly the transmission and emission (spectro)photometry techniques. During a transit, some of the starlight passes through the planetary atmosphere and, depending on the atmospheric extent, temperature, and composition, wavelength-dependent variations are seen in the amount of absorbed flux. Thus, from multi-wavelength transit light curves, a transmission spectrum of the upper atmosphere at the day-night terminator region can be obtained (e.g. \citealt{seager2000}, \citealt{charbonneau2002}). At the opposite conjunction, when the planet passes behind the star during a secondary eclipse (occultation), one can measure the flux drop caused by the elimination of the flux component originating from the dayside of the planet (e.g. \citealt{deming2005}, \citealt{charbonneau2005}). Using this technique at different wavelengths allows to probe the emission spectrum of the planet's dayside, from which insights on its atmospheric composition and vertical pressure-temperature ($P-T$) profile can be gained.\\
\indent
A broad diversity of transmission spectra has been found across the population of close-in transiting gas giant exoplanets (see e.g. \citealt{sing2016}, \citealt{fu2017}, \citealt{tsiaras2017}). Some planets have sufficiently clear atmospheres to allow detections of atomic and molecular species, in particular Na (e.g. \citealt{nikolov14h1}), K (e.g. \citealt{sing2011pot}), and $\mathrm{H_{2}O}$ (e.g. \citealt{deming2013}), while others appear to contain high-altitude clouds or hazes that completely mute the spectral features of the atmospheric components (see e.g. \citealt{gibson2013}, \citealt{line2013h12}, \citealt{lendl2016}). Even when detected, atmospheric spectral signatures are often less pronounced than predicted by theoretical models of clear atmospheres with solar abundances (e.g. \citealt{deming2013}, \citealt{madhureview}), suggesting that an extra opacity source is still present, at some level, in otherwise predominantly cloud-free atmospheres. This picture is supported by evidences for high-altitude atmospheric hazes reported for several hot Jupiters (see e.g. \citealt{nikolov2015}, \citealt{sing2015}, \citealt{sing2016}), based on their Rayleigh or Mie scattering signature in the planets' transmission spectra.\\ 
\indent 
On the emission side, most observations gathered so far seem to indicate atmospheric vertical pressure-temperature profiles without significant thermal inversions. Temperature inversions were predicted by early theoretical studies of highly irradiated giant planets, that suggested two classes of hot Jupiters based on their degree of irradiation (e.g. \citealt{hubeny2003}, \citealt{fortney2008}); the hotter class was predicted to host thermal inversions in their atmospheres due to strong absorption of incident UV/visible irradiation at high altitude by high-temperature absorbers, such as gaseous TiO and VO (commonly found in low-mass stars and brown dwarfs), while cooler atmospheres were expected to be devoid of thermal inversions due to the condensation of these absorbing compounds. Thermal inversions have been previously claimed for several hot Jupiters based on \textit{Spitzer} observations (e.g. \citealt{knutson2008}, \citealt{machalek2008}, \citealt{knutson2009}), but these detections have been seriously called into question since then (e.g. \citealt{hansen2014}, \citealt{diamond2014}, \citealt{schwarz2015}). It has been suggested that TiO and VO may not remain suspended in the upper atmospheres of hot Jupiters due to cold-trapping, that would occur either deeper in the dayside atmosphere or on the cooler nightside, and would cause their condensation and downward drag by gravitational settling (e.g. \citealt{spiegel2009}). Inversion-causing compounds may also be photodissociated by high chromospheric emission from the host star, so that the formation of inversions may be affected by stellar activity (e.g. \citealt{knutsonact}). Another important factor is the atmospheric chemistry; for example, the atmospheric carbon-to-oxygen ratio (C/O) can control the abundance of TiO/VO, with a C/O $\geq$ 1 causing substantial depletion of TiO/VO, most available oxygen being taken up by CO molecules in this case thus leaving no oxygen for gaseous TiO/VO (\citealt{madhu2012}). Alternatively, the apparently isothermal emission spectra observed for many hot Jupiters (e.g. \citealt{hansen2014}) may result from the presence of high-altitude cloud decks in their dayside atmospheres that could prevent some thermal inversions from being detected (e.g. \citealt{singW12}). Nevertheless, hottest planets are still the best candidates to look for thermal inversions. Indeed, the planets showing the strongest evidence to date for temperature inversions are WASP-33\,b (\citealt{haynes2015}, \citealt{Nugroho2017}) and WASP-121\,b (\citealt{evans2017}), which are among the most highly irradiated hot Jupiters currently known. In this work, we study the atmospheric properties of another ultra-hot gas giant, \hbox{WASP-103\,b} (\citealt{gillon14}, hereafter \citetalias{gillon14}).\\
\indent
This extreme hot Jupiter, discovered by the WASP Collaboration (\citealt{pollacco2006}, \citealt{andrew2007}, \citealt{hellier2011}), has a mass of $\sim$1.5 $M_{\mathrm{Jup}}$, an inflated radius of $\sim$1.6 $R_{\mathrm{Jup}}$, and is in an ultra-short-period orbit ($\sim$22.2hrs) around a relatively bright (\hbox{$V$ = 12.1}, \hbox{$K$ = 10.8)} F8V star (\citetalias{gillon14}). With an incident stellar flux of \hbox{$\sim$$9.1\times10^{9}$ erg $\mathrm{s}^{-1} \mathrm{cm}^{-2}$} (\hbox{$\sim$$9.1\times10^{6}$ W $\mathrm{m}^{-2}$}), it is one of the most highly irradiated hot Jupiters known to date. Assuming a null Bond albedo, it is heated to an equilibrium temperature close to \hbox{2500 K}. These properties make \hbox{WASP-103\,b} an exquisite target for atmospheric characterization. Another interesting fact about this planet is that its orbital semi-major axis is also only $\sim$1.16 times larger than its Roche limit, meaning that the planet might be close to tidal disruption. \hbox{WASP-103\,b} is thus also a favorable object for studying the atmospheric properties of hot Jupiters in the last stages of their evolution.\\
\indent
\citeauthor{southworth15} ({\color{blue}2015}, hereafter \citetalias{southworth15}) published high-precision follow-up transit photometry of WASP-103\,b in several broad-band optical filters, which they used to refine the physical and orbital parameters of the system. They also detected a slope in the resulting broad-band transmission spectrum, larger values of the effective planetary radius being obtained at bluer wavelengths, which they found to be too steep to be mainly caused by Rayleigh scattering in the planetary atmosphere. Subsequent to their study, a previously unresolved faint star was found via lucky imaging by \citeauthor{wollert2015} ({\color{blue}2015}, hereafter \citetalias{wollert2015}) at an angular separation of only 0.24$\arcsec$ from WASP-103. This object, which was also recently imaged by \citeauthor{ngo16} ({\color{blue}2016}, hereafter \citetalias{ngo16}), is significantly redder than WASP-103 and may be either gravitationally bound or simply aligned along the line of sight. Contamination from this redder star, if not accounted for, is expected to produce a blueward slope in the transmission spectrum of WASP-103\,b, the transit signal being more strongly diluted at longer wavelengths than at shorter ones. This led \citeauthor{southworth16} ({\color{blue}2016}, hereafter \citetalias{southworth16}) to publish a reanalysis of the data presented in \citetalias{southworth15}, accounting for the presence of the contaminating star. They found the inclusion of contaminating light from the faint star in their analysis to have no significant effect on the derived system physical properties. They also reported a corrected broad-band transmission spectrum showing, instead of a steep slope, a minimum effective planetary radius around 760 nm and increasing values towards both bluer and redder wavelengths. This ``V-shape" is not well reproduced by existing theoretical models of transmission spectra. Very recently, \cite{lendl2017} reported an optical transmission spectrum of WASP-103\,b obtained at medium spectral resolution between 550 and \hbox{960 nm} using Gemini/GMOS. While they found signs of strong Na and K absorption, they did not observe any evidence for the V-shape pattern reported by \citetalias{southworth16}.\\
\indent
Recently, \citeauthor{cartier17} ({\color{blue}2017}, hereafter \citetalias{cartier17}) presented near-infrared occultation spectrophotometry of WASP-103\,b obtained using the \textit{Hubble Space Telescope}/Wide Field Camera 3. After correction for flux contamination from the nearby star, they found the dayside emission spectrum of WASP-103\,b to be indistinguishable from isothermal from 1.1 to 1.7 $\mu$m. They noted that several atmospheric models, besides an isothermal one, can result in an apparently isothermal emission spectrum across this wavelength range, for example an atmosphere with a thermal inversion layer just above the layer probed by their observations, an atmosphere with a monotonically decreasing temperature-pressure profile and a C/O>1, or an atmosphere harboring clouds or hazes at high altitude. This highlights the need for additional eclipse observations at other wavelengths to help differentiate between these potential atmospheric scenarios.\\
\indent
To improve the atmospheric characterization of \hbox{WASP-103\,b}, we carried out an intense ground-based photometric monitoring of its occultations, with the aim of probing its dayside emission spectrum in the $z'$ (0.9 $\mu$m) and $K_{\mathrm{S}}$ (2.1 $\mu$m) bands. We complemented the data acquired in the frame of this program with some additional transit photometry and radial velocity (RV) measurements, combined all these new observations with the data previously published in \citetalias{gillon14} and \citetalias{southworth15}, and performed a global analysis of the resulting extensive dataset, taking into account the contamination from the faint star. The Gemini/GMOS transmission and HST/WFC3 emission data were published during the final stages of the preparation of this manuscript, so we did not include them in our global analysis but we discuss these measurements along with our results in the scientific discussion. The paper is organized as follows. The new observations and their reduction are described in Section \ref{obs}, as well as the archival data used in our global analysis. In Section \ref{analysis}, we present our detailed data analysis and results. We discuss these results in Section \ref{discuss}, before concluding in Section \ref{concl}.

\vspace{-0.2cm}

\section{Observations and Data Reduction}
\label{obs}

%\vspace{0.3cm}
\subsection{New data}

Between May 2014 and July 2015, we gathered a total of nineteen eclipse light curves of WASP-103\,b. Sixteen of these light curves were acquired during occultations of the planet and three during transits. This follow-up photometry was obtained using three different instruments: the 0.6m TRAPPIST robotic telescope and the EulerCam CCD camera on the 1.2m \textit{Euler}-Swiss telescope, both located at ESO La Silla Observatory (Chile), as well as the WIRCam near-infrared imager on the 3.6m Canada-France-Hawaii Telescope (CFHT) at Mauna Kea Observatory (Hawaii). We complemented this dataset with five new RV measurements obtained between Sept. 2013 and Sept. 2014 with the CORALIE spectrograph mounted on the \textit{Euler} telescope. The follow-up light curves are summarized in the upper part of Table \ref{tablelog}, while the RVs are presented in Table \ref{tablervs}. We describe these new data in the sections below.

\begin{table*}
\centering
\caption{Summary of follow-up photometry obtained for WASP-103. For each light curve, this table shows the night of acquisition (UT), the used instrument, the eclipse type, the filter ($BB$=blue-blocking) and exposure time, the number of data points, the selected baseline function, the standard deviation of the best-fit residuals (unbinned and binned per intervals of 2 min), and the deduced values for $\beta_{w}$, $\beta_{r}$ and $CF=\beta_{w} \times \beta_{r}$ (see Section \ref{analysis_method} for details). For the baseline function, p($\epsilon^{N}$) denotes, respectively, a $N$-order polynomial function of time ($\epsilon=t$), airmass ($\epsilon=a$), PSF full-width at half maximum ($\epsilon=f$) or radius for the WIRCam data ($\epsilon=r$), background ($\epsilon=b$), and $x$ and $y$ positions ($\epsilon=xy$). $o$ denotes an offset fixed at the time of the meridian flip.}
\vspace{-0.1cm}
\begin{tabular}{cccccccccccc}
  \hline
  \hline 
 Date&Instrument&Eclipse&Filter&$T_{\mathrm{exp}}$&$N_{\mathrm{p}}$&Baseline&$\sigma$&$\sigma_{120\mathrm{s}}$& $\beta_{w}$&$\beta_{r}$&$CF$\\
 (UT) & & type & & (s) & &function&(\%)&(\%)& & &\\
  \hline
  New data & & & & & & & & & & &\\
  \hline
  2014 May 08-09&TRAPPIST&Occultation&$z'$&55&338&p($t^{1}$) + $o$&0.25&0.19&1.10&1.08&1.18\\
  2014 May 20-20&WIRCam&Occultation&$K_{\mathrm{S}}$&5&1083&p($t^{2}$+$r^{1}$+$b^{1}$)&0.32&0.12&1.56&1.00&1.57\\
  2014 June 10-11&EulerCam&Transit&$r'$&80&140&p($t^{1}$+$b^{1}$)&0.18&0.18&1.89&1.00&1.89\\
  2014 June 16-17&TRAPPIST&Occultation&$z'$&50&341&p($t^{1}$) + $o$&0.25&0.18&1.08&1.07&1.16\\
  2014 June 29-30&EulerCam&Occultation&$z'$&100&130&p($t^{1}$+$f^{1}$)&0.10&0.10&1.20&1.00&1.20\\
  2014 July 05-06&EulerCam&Transit&$r'$&80&150&p($t^{1}$)&0.09&0.09&1.16&1.49&1.72\\
  2014 July 11-12&EulerCam&Occultation&$z'$&100&65&p($t^{1}$+$b^{1}$)&0.09&0.09&1.18&1.10&1.30\\
  2014 July 12-13&TRAPPIST&Occultation&$z'$&48&281&p($t^{1}$) + $o$&0.26&0.18&1.05&1.91&2.01\\
  2014 July 25-26&TRAPPIST&Occultation&$z'$&48&308&p($t^{1}$) + $o$&0.23&0.16&1.00&1.10&1.10\\
  2015 Apr. 02-03&TRAPPIST&Transit&$BB$&8&910&p($a^{1}$) + $o$&0.35&0.14&0.87&1.08&0.95\\
  2015 May 17-18&TRAPPIST&Occultation&$z'$&36&328&p($t^{1}$) + $o$&0.32&0.23&1.28&1.22&1.56\\
  2015 May 30-31&TRAPPIST&Occultation&$z'$&36&364&p($a^{1}$) + $o$&0.36&0.26&1.32&1.00&1.32\\
  2015 June 11-12&EulerCam&Occultation&$z'$&100&139&p($t^{2}$+$f^{1}$+$b^{1}$)&0.14&0.14&1.56&2.10&3.27\\
  2015 June 12-13&TRAPPIST&Occultation&$z'$&40&315&p($a^{1}$) + $o$&0.36&0.26&1.43&1.00&1.43\\
  2015 July 07-08&TRAPPIST&Occultation&$z'$&46&365&p($t^{1}$) + $o$&0.30&0.22&1.29&1.51&1.94\\
  2015 July 07-08&EulerCam&Occultation&$z'$&100&126&p($a^{2}$+$f^{1}$+$b^{1}$)&0.11&0.11&1.27&1.42&1.82\\
  2015 July 19-20&TRAPPIST&Occultation&$z'$&46&327&p($t^{1}$) + $o$&0.35&0.24&1.37&1.17&1.60\\
  2015 July 19-20&EulerCam&Occultation&$z'$&100&134&p($t^{1}$+$f^{1}$+$b^{1}$)&0.10&0.10&1.13&1.22&1.39\\
  2015 July 20-21&EulerCam&Occultation&$z'$&60&210&p($t^{1}$+$f^{1}$)&0.10&0.08&1.10&1.37&1.51\\
  \hline
  Archival data & & & & & & & & & & &\\
  \hline
  2013 June 15-16&TRAPPIST&Transit&$BB$&7&802&p($t^{1}$+$f^{1}$) + $o$&0.32&0.13&0.69&1.48&1.02\\ 
  2013 June 28-29&EulerCam&Transit&$r'$&120&103&p($t^{1}$+$f^{1}$+$b^{2}$+$xy^{1}$)&0.09&0.09&1.43&1.08&1.54\\ 
  2013 July 11-12&EulerCam&Transit&$r'$&80&105&p($t^{1}$+$b^{1}$)&0.12&0.12&1.59&1.23&1.96\\ 
  2013 July 23-24&TRAPPIST&Transit&$BB$&9&941&p($t^{2}$) + $o$&0.48&0.19&1.23&1.22&1.50\\ 
  2013 Aug. 04-05&TRAPPIST&Transit&$BB$&10&935&p($t^{2}$) + $o$&0.29&0.13&0.80&1.51&1.21\\ 
  2014 Apr. 19-20&DFOSC&Transit&$R$&100-105&134&p($t^{1}$)&0.07&0.07&1.05&2.53&2.65\\ 
  2014 May 01-02&DFOSC&Transit&$I$&110-130&113&p($t^{1}$)&0.08&0.08&1.06&1.14&1.21\\ 
  2014 June 08-09&DFOSC&Transit&$R$&100-130&130&p($t^{1}$)&0.10&0.10&1.03&1.41&1.45\\ 
  2014 June 22-23&DFOSC&Transit&$R$&50-120&195&p($t^{1}$)&0.13&0.13&1.05&2.62&2.75\\ 
  2014 June 23-24&DFOSC&Transit&$R$&100&112&p($t^{1}$)&0.06&0.06&1.01&1.44&1.45\\ 
  2014 July 05-06&DFOSC&Transit&$R$&100&118&p($t^{1}$)&0.06&0.06&1.01&1.81&1.83\\ 
  2014 July 05-06&GROND&Transit&$g'$&100-120&122&p($t^{2}$)&0.12&0.12&1.03&1.44&1.49\\ 
  2014 July 05-06&GROND&Transit&$r'$&100-120&125&p($t^{2}$)&0.07&0.07&1.01&2.20&2.22\\ 
  2014 July 05-06&GROND&Transit&$i'$&100-120&119&p($t^{2}$)&0.08&0.08&1.04&1.42&1.48\\ 
  2014 July 05-06&GROND&Transit&$z'$&100-120&121&p($t^{2}$)&0.11&0.11&1.03&2.49&2.58\\ 
  2014 July 17-18&DFOSC&Transit&$R$&90-110&139&p($t^{1}$)&0.07&0.07&1.00&1.21&1.22\\ 
  2014 July 18-19&DFOSC&Transit&$R$&60-110&181&p($t^{2}$)&0.06&0.05&1.02&1.19&1.21\\
  2014 July 18-19&GROND&Transit&$g'$&98-108&126&p($t^{2}$)&0.08&0.08&1.02&1.44&1.48\\ 
  2014 July 18-19&GROND&Transit&$r'$&98-108&143&p($t^{2}$)&0.06&0.06&1.04&1.64&1.70\\ 
  2014 July 18-19&GROND&Transit&$i'$&98-108&142&p($t^{2}$)&0.09&0.09&1.05&1.66&1.74\\ 
  2014 July 18-19&GROND&Transit&$z'$&98-108&144&p($t^{2}$)&0.09&0.09&1.01&1.13&1.14\\ 
  2014 Aug. 12-13&CASLEO&Transit&$R$&90-120&129&p($t^{2}$)&0.15&0.15&1.04&1.51&1.58\\ 
  \hline
  \hline
\end{tabular}
\label{tablelog}
\vspace{-0.3cm}
\end{table*}

\vspace{-0.2cm}
\subsubsection{TRAPPIST eclipse photometry}

We observed one transit and nine occultations of \hbox{WASP-103\,b} using the 0.6m TRAPPIST robotic telescope and its thermoelectrically-cooled 2K$\times$2K CCD (field of view of 22\arcmin$\times$22\arcmin, plate scale of 0.65\arcsec/pixel). For details of TRAPPIST, see \cite{gillontrap} and \cite{jehintrap}. The transit was observed in a blue-blocking ($BB$) filter that has a transmittance $>$90\% from 500 nm to beyond 1000 nm (effective wavelength = 696.8 nm), with each frame exposed for 8s. The occultations were acquired through a Sloan-$z'$ filter (effective wavelength = 895.5 nm), with exposure times between 36s and 55s. Throughout observations, the telescope was kept in focus and the positions of the stars on the chip were retained on the same few pixels, thanks to a ``software guiding" system that regularly derives an astrometric solution on the images and sends pointing corrections to the mount when needed.\\
\indent
After bias, dark, and flat-field corrections, stellar fluxes were extracted from the images using the \textsc{iraf/daophot}\footnote{\textsc{iraf} is distributed by the National Optical Astronomy Observatory, which is operated by the Association of Universities for Research in Astronomy, Inc., under cooperative agreement with the National Science Foundation.} aperture photometry software (\citealt{stetson}). For each observation, a careful selection of both the photometric aperture size and of stable reference stars having a brightness similar to \hbox{WASP-103} was performed to obtain optimal differential photometry. The resulting light curves are shown in Figs. \ref{phot_occz} (raw occultation light curves), \ref{phoc_occz} (detrended occultation light curves, see Section \ref{analysis_method} for details about the modeling), and \ref{transits1} (raw transit light curve).

\vspace{-0.2cm}
\subsubsection{Euler/EulerCam eclipse photometry}

Two transit and six occultation light curves of WASP-103\,b were obtained with EulerCam, the imager of the 1.2m \textit{Euler}-Swiss telescope. EulerCam is a nitrogen-cooled 4K$\times$4K CCD camera with a field of view of 15.68\arcmin$\times$15.73\arcmin$\:$at a plate scale of 0.215\arcsec/pixel. The transits were observed in a Gunn-$r'$ filter (effective wavelength = 664.1 nm) with an exposure time of 80s. The occultations were acquired through a Gunn-$z'$ filter (effective wavelength = 912.3 nm), with exposure times between 60s and 100s. A slight defocus was applied to the telescope to optimize the observation efficiency and to minimize pixel-to-pixel effects. This resulted in stellar PSFs with a typical FWHM between 1.1\arcsec$\:$and 2.5\arcsec. Here too, the positions of the stars on the detector were kept within a box of a few pixels throughout the observations, thanks to the ``Absolute Tracking'' system of EulerCam that matches the point sources in each image with a catalog and adjusts the telescope pointing between exposures when needed. The reduction procedure used to extract the eclipse light curves was similar to that performed on TRAPPIST data. The resulting light curves are shown in Figs. \ref{phot_occz} (raw occultation light curves), \ref{phoc_occz} (detrended occultation light curves, see Section \ref{analysis_method} for details about the modeling), and \ref{transits1} (raw transit light curves). Further details of the EulerCam instrument and data reduction procedures can be found in \cite{lendleuler}.

\vspace{-0.2cm}
\subsubsection{CFHT/WIRCam occultation photometry}

We observed one occultation of WASP-103\,b with the Wide-field InfraRed Camera (WIRCam, \citealt{wircam}) on the 3.6m Canada-France-Hawaii Telescope. WIRCam consists of four 2K$\times$2K HgCdTe HAWAII-2RG arrays, arranged in a 2$\times$2 mosaic. The instrument has a total field of view of 20.5\arcmin$\times$20.5\arcmin$\:$(with gaps of 45\arcsec$\:$between adjacent chips) at a scale of 0.3\arcsec/pixel. We used the $K_{\mathrm{S}}$ broad-band filter, which has a bandwidth of 0.325 $\mu$m centered at \hbox{2.146 $\mu$m}. The observations took place on 2014 May 20 from 06:50 to 12:40 UT, covering the 2.6hrs long predicted occultation (assuming a circular orbit) together with 3.2hrs of out-of-eclipse observations. Conditions were photometric, with a median seeing of 0.6\arcsec, and airmass decreased from 2.4 to 1.09 during the run. The data were gathered in staring mode (\citealt{devost}), with the target and reference stars observed continuously for several hours on the same pixels without any dithering. This mode has been used by other authors for similar observations with WIRCam (see, e.g., \citealt{wang} and \citealt{croll2015}) and has proven to yield an optimal photometric precision. The pointing was carefully selected to ensure that the target and reference stars did not fall near bad pixels, as well as to maximize the number of suitable reference stars located on the same WIRCam chip as WASP-103 (see below). 
The scientific sequence consisted of 1092 exposures, each 5s and read out with correlated double sampling\footnote{A CDS image is constructed by subtracting a first read of the array, done immediately after reset, from a second read of the array, performed at the end of the exposure.} (CDS). A defocus of 2mm was applied to the telescope, in order to reduce the impact of imperfect flat-fielding, inter- and intra-pixel variations on the photometry, as well as to keep the counts of the target and reference stars in the regime where detector non-linearity is minimized (linearity to within 1\% below $\sim$10 kADU)\footnote{\burl{www.cfht.hawaii.edu/Instruments/Imaging/WIRCam/WIRCamNonlinearity.html}\label{note}}. This resulted in a ring-shaped PSF with a radius of $\sim$4.5\arcsec ($\sim$15 pixels). A short set of dithered (ten offset positions) in-focus images was taken before and after the scientific sequence in order to construct a sky flat (see below).\\
\indent
The data were reduced independently of the traditional WIRCam \`{}I\`{}iwi pipeline\footnote{\burl{http://www.cfht.hawaii.edu/Instruments/Imaging/WIRCam/IiwiVersion2Doc.html}}, following the prescription of \cite{croll2015} for the reduction of WIRCam staring mode data. We refer the reader to that paper for a detailed description of the reduction procedure and give here an outline of the main steps. The frames from each detector were reduced separately. In each image, the pixels with CDS values above 36 kADU were flagged as saturated. The data were corrected for the small non-linearity\footref{note} present below \hbox{10 kADU}, following the iterative approach of \cite{vacca} for applying a non-linearity correction to CDS images. Each frame was dark subtracted and divided by a sky flat, which was created by taking the median stack of the dithered images acquired before and after the scientific sequence. A bad pixel map was constructed from this sky flat, where pixels flagged as bad were those that deviate by more than 2\% from the median of the array. Finally, in each image, bad and saturated pixels had their value replaced by the median value of the adjacent pixels (provided that they were not themselves bad or saturated).

\begin{figure}
\centering     
\vspace{0.1cm}
\includegraphics[scale=0.43] {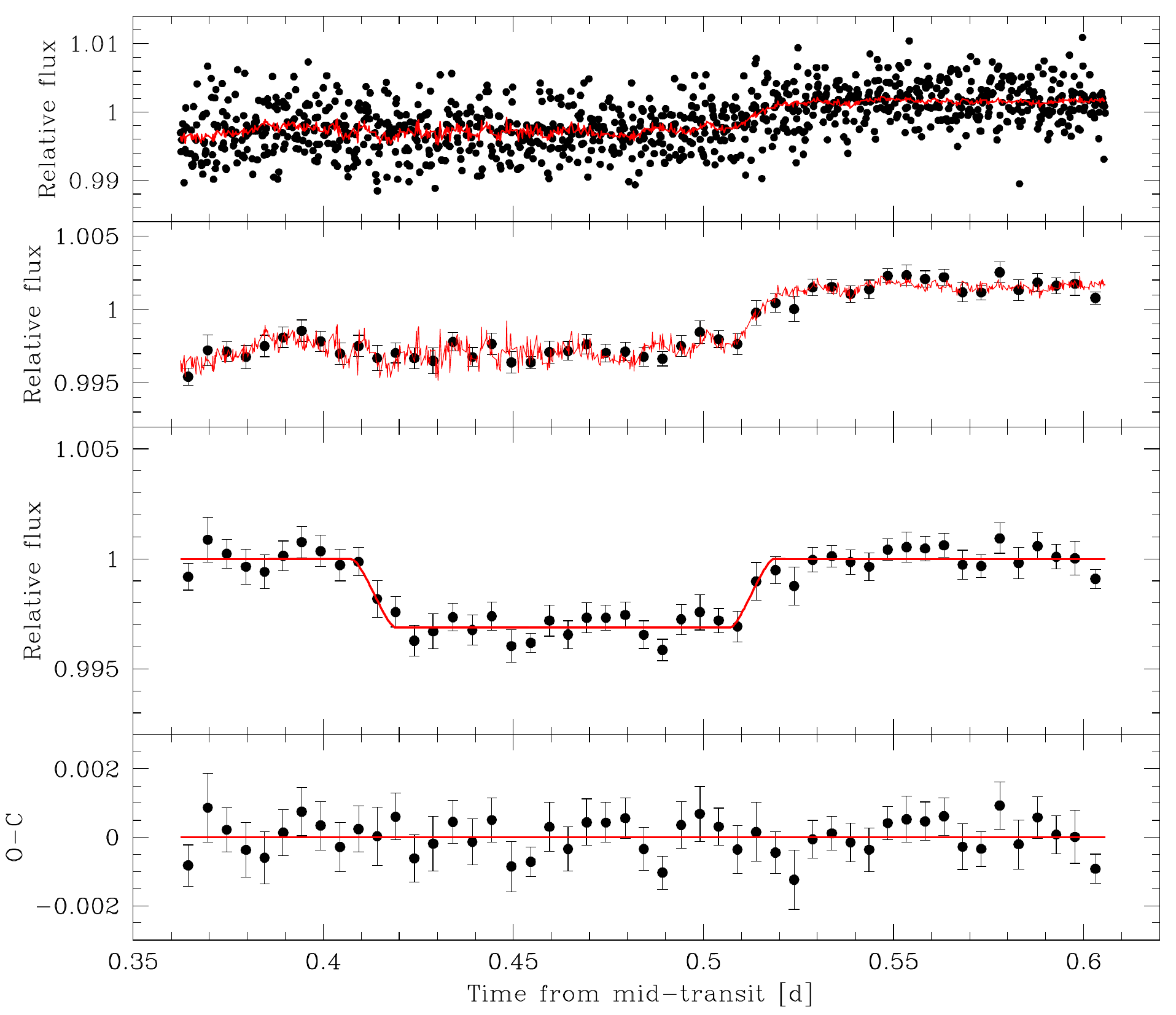} 
\vspace{-0.5cm}
\caption{Occultation photometry obtained with CFHT/WIRCam in the $K_{\mathrm{S}}$-band. \textit{Top panel:} the raw, unbinned light curve, together with the best-fit full (photometric baseline $\times$ occultation in the $K_{\mathrm{S}}$-band) model (overplotted in red, see Section \ref{analysis_method} for details about the modeling). \textit{Second panel:} same as the top panel except that the data are binned in 7.2min bins. \textit{Third panel:} binned light curve divided by the best-fit photometric baseline model. The best-fit occultation model in the $K_{\mathrm{S}}$-band is overplotted in red. \textit{Bottom panel:} best-fit residuals. The RMS of the residuals is 524 ppm (7.2min bins). \textit{General note:} the data and models are not corrected for the dilution by the nearby star here.}     
\label{lcs_Ks}
\vspace{-0.25cm}
\end{figure}

\indent
Aperture photometry was performed for the target and reference stars using \textsc{iraf/daophot}. Apertures were centered using intensity-weighted centroids. We tested a set of constant apertures, as well as apertures that varied from image to image as a function of the mean radius of the ring-shaped stellar PSFs. The best result was obtained with a variable aperture of 1.1 times the mean radius. Using a variable photometric aperture, rather than a fixed one, allowed us to account for the varying atmospheric conditions, as well as to find for each individual image a balance between choosing a small aperture to minimize the sky noise and a large aperture to encompass all the stellar light. For other examples of exoplanet near-infrared photometry extracted using a variable aperture size, see e.g. \cite{lendl2013} or \cite{zhouw19}. The sky annulus was kept constant for all images, with an inner radius of 30 pixels and an outer radius of 50 pixels. Differential photometry of WASP-103 was finally obtained. We tested all possible combinations of stable reference stars having a brightness similar to the target. We found the best photometry using eight reference stars located on the same WIRCam detector as WASP-103. The resulting light curve is shown in Fig. \ref{lcs_Ks}.

\vspace{-0.2cm}
\subsubsection{Euler/CORALIE radial velocities}

\begin{figure}
\centering     
%\vspace{-0.2cm}
\includegraphics[scale=0.43] {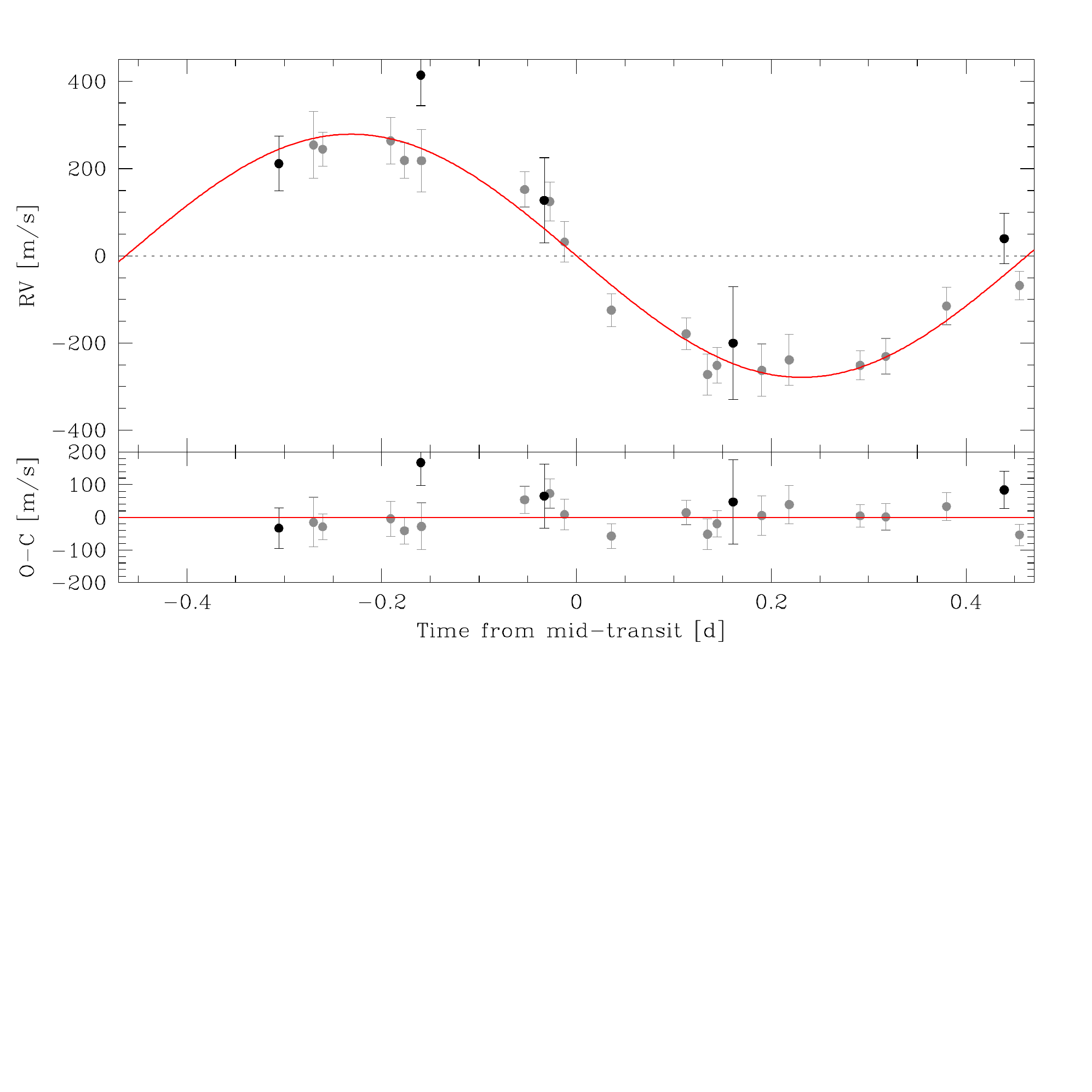} 
\vspace{-3.8cm}
\caption{\textit{Top:} Euler/CORALIE RV measurements period-folded on the best-fit transit ephemeris from our global MCMC analysis (see Section \ref{analysis_method}), with the best-fit Keplerian model overplotted in red. The data published in \citetalias{gillon14} are plotted in grey, while our new measurements are plotted in black. \textit{Bottom:} corresponding residuals.}
%\vspace{-0.1cm}     
\label{rvs}
\end{figure}

Five new spectroscopic measurements of WASP-103 were gathered with the CORALIE spectrograph mounted on the \textit{Euler} telescope (\citealt{queloz2000}). The spectra, all obtained with an exposure time of 30min, were processed with the CORALIE standard data reduction pipeline (\citealt{baranne96}). RVs were then computed from the spectra by weighted cross-correlation (\citealt{pepe2002}), using a numerical G2-spectral template that provides optimal precisions for late-F to early-K dwarfs. These RVs are presented in Table \ref{tablervs} and shown in Fig. \ref{rvs}. The cross-correlation function (CCF) FWHM and bisector span (BS, \citealt{queloz2001}) values are also given in Table \ref{tablervs}.

\begin{table}
\centering
\caption{New CORALIE RVs for WASP-103. The last two columns give the cross-correlation function FWHM and bisector span values, respectively.}
\label{tablervs}
\vspace{-0.1cm}
\begin{tabular}{ccccc}
\hline
BJD & RV & $\mathrm{\sigma}_{\mathrm{RV}}$ & FWHM & BS\\
- 2 450 000 & (km $\mathrm{s}^{-1}$) & (km $\mathrm{s}^{-1}$) & (km $\mathrm{s}^{-1}$) & (km $\mathrm{s}^{-1}$)\\
\hline
6537.506214 & -42.19467 & 0.12930 & 15.47857 & 0.04812\\
6837.661259 & -41.95497 & 0.05763 & 14.90643 & 0.07458\\
6852.650727 & -41.78295 & 0.06256 & 14.92542 & 0.26543\\
6880.562742 & -41.57979 & 0.07013 & 15.07407 & 0.45440\\
6920.487957 & -41.86693 & 0.09804 & 14.64620 & -0.40181\\
\hline
\end{tabular}
%\vspace{-0.2cm}
\end{table}

\subsection{Archival data}
\vspace{0.1cm}

We also included in our global analysis the data previously published in \citetalias{gillon14} and \citetalias{southworth15}:

\begin{itemize}
\item three TRAPPIST transit light curves (blue-blocking \hbox{filter)};
\item two Euler/EulerCam transit light curves (Gunn-$r'$ filter);
\item eight transit light curves gathered with the DFOSC (Danish Faint Object Spectrograph and Camera) instrument on the 1.54m Danish telescope located at ESO La Silla Observatory (Bessel $R$ and $I$ filters);
\item eight transit light curves obtained using the GROND (Gamma-Ray Burst Optical/Near-Infrared Detector) instrument on the 2.2m MPG/ESO telescope (Sloan-$g'$, -$r'$, -$i'$, and -$z'$ filters);
\item one transit light curve acquired with the 2.15m telescope located at the CASLEO (Complejo Astronomico El Leoncito) Observatory (Johnson-Cousins $R$ filter);
\item eighteen CORALIE RVs.
\end{itemize}

\noindent
We refer the reader to \citetalias{gillon14} and \citetalias{southworth15} for more details about these data. The archival light curves are summarized in the lower part of Table \ref{tablelog} and shown in Figs. \ref{transits1} (TRAPPIST and Euler/EulerCam), \ref{transits2} (Danish/DFOSC), and \ref{transits3} (2.2m/GROND and CASLEO/2.15m), while the RVs are shown in Fig. \ref{rvs}.

\vspace{0.1cm}
\section{Data Analysis}
\label{analysis}

\vspace{0.2cm}
\subsection{Contamination from the nearby star}
\label{contam}

\vspace{0.2cm}

\citetalias{wollert2015} reported the detection of a stellar object located 0.242$\pm$0.016$\arcsec$ from WASP-103, fainter by $\Delta i'=3.11\pm0.46$ and $\Delta z'=2.59\pm0.35$. \citetalias{ngo16} presented further observations of this nearby star in the near-infrared and found magnitude differences with WASP-103 of $\Delta J=2.427\pm0.030$, $\Delta H=2.2165\pm0.0098$, and $\Delta K_{\mathrm{S}}=1.965\pm0.019$. The astrometric measurements from these two studies are inconclusive as to whether this star is gravitationally bound to the planetary system or not. Due to the very small angular separation of WASP-103 and the nearby object, both stars are contained in all photometric apertures used to extract the eclipse light curves that we included in our global analysis (new and archival data). Although a detailed characterization of the nearby star is beyond the scope of this work, we must estimate the dilution correction factor $(F_{\mathrm{W103}}+F_{\mathrm{cont}})/F_{\mathrm{W103}}=1+F_{\mathrm{cont}}/F_{\mathrm{W103}}$ (where $F_{\mathrm{cont}}/F_{\mathrm{W103}}$ is the contaminant-to-target flux ratio) for each of the observed passbands, in order to account for this contamination in our data analysis.

\begin{table}
\centering
\vspace{0.1cm}
\begin{tabular}{ccc}
\hline
Filter & WASP-103 & Contaminant\\
\hline
$J$ & 11.210 $\pm$ 0.029 & 13.637 $\pm$ 0.053\\
$H$ & 10.993 $\pm$ 0.032 & 13.209 $\pm$ 0.039\\
$K_{\mathrm{S}}$ & 10.932 $\pm$ 0.023 & 12.897 $\pm$ 0.037\\
\hline
\end{tabular}
\caption{Individual apparent magnitudes of WASP-103 and the nearby star in the $J$, $H$, and $K_{\mathrm{S}}$ bands.}
\label{magnitude}
\vspace{0.3cm}
\end{table}

\indent 
To this end, we first derived the individual apparent magnitudes of WASP-103 and the nearby star in the $J$, $H$, and $K_{\mathrm{S}}$ bands based on their combined 2MASS magnitudes (\citealt{2mass}) and the magnitude differences reported by \citetalias{ngo16} in these bands. The resulting apparent magnitudes are given in Table \ref{magnitude}. While the $J-H$, $H-K_{\mathrm{S}}$, and $J-K_{\mathrm{S}}$ color indices of WASP-103 agree well with a F8V star, the colors of the nearby star suggest a spectral type comprised between K1 and M4 if it is on the main sequence (\citealt{colortype}). We computed the flux ratios $F_{\mathrm{cont}}/F_{\mathrm{W103}}$ in the passbands of interest assuming each of these two spectral types for the contaminant. We used for this purpose PHOENIX model spectra (\citealt{husser13}) of WASP-103 and the nearby star that we integrated over the passbands of interest. The flux ratio in a given passband can be expressed as:
\begin{equation}
\frac{F_{\mathrm{cont}}}{F_{\mathrm{W103}}} = f^{2} \frac{M_{\mathrm{cont}}}{M_{\mathrm{W103}}}
\end{equation}
where $M_{\mathrm{W103}}$ and $M_{\mathrm{cont}}$ are the band-integrated model fluxes of WASP-103 and the contaminant, respectively, and $f$ is a geometric factor defined as:
\begin{equation}
f=\frac{R_{\mathrm{cont}}}{R_{\mathrm{W103}}} \frac{d_{\mathrm{W103}}}{d_{\mathrm{cont}}}
\end{equation}
with $R_{\mathrm{W103}}$ (resp. $R_{\mathrm{cont}}$) and $d_{\mathrm{W103}}$ (resp. $d_{\mathrm{cont}}$) denoting, respectively, the radius and the distance of WASP-103 (resp. the contaminant). For WASP-103, we used a model spectrum interpolated to the effective temperature $T_{\mathrm{eff}}$, surface gravity log $g_{\star}$, and metallicity [Fe/H] reported in \citetalias{gillon14}. For the nearby star, two different model spectra, of typical K1V ($T_{\mathrm{eff}}$=5100K, log $g_{\star}$=4.5, [Fe/H]=0.0) and M4V ($T_{\mathrm{eff}}$=3200K, log $g_{\star}$=5.0, [Fe/H]=0.0) stars, were used. In each of these two cases, we calculated first the wavelength-independent factor $f$ by comparing the ratio of the model fluxes $M_{\mathrm{cont}}/M_{\mathrm{W103}}$ integrated over the $J$-band with the flux ratio $F_{\mathrm{cont}}/F_{\mathrm{W103}}$ measured in this band by \citetalias{ngo16}. The flux ratios in the passbands of interest were then computed using Equation (1). The final value for the flux ratio in each passband was taken as the average of the two values obtained using the K1V and M4V model spectra for the contaminating star, with an error bar large enough to encompass both values. The resulting flux ratios are given in Table \ref{dilution}. The values derived in the $i'$ and $z'$ bands are consistent with the measurements from \citetalias{wollert2015} in these two bands ($0.0622\pm0.0250$ and $0.0969\pm0.0302$, respectively), but more precise (in the $z'$-band). In the $K_{\mathrm{S}}$-band, we obtained a flux ratio of $0.1496\pm0.0178$, consistent with the value of $0.1637\pm0.0029$ measured by \citetalias{ngo16}, but significantly less precise. For this passband, we thus used the measurement from \citetalias{ngo16}.\\ 
\indent
In their reanalysis, \citetalias{southworth16} also estimated the contamination from the nearby star in the passbands observed in \citetalias{southworth15}, using the near-infrared magnitude differences reported by \citetalias{ngo16}.  
They obtained values in good agreement with ours, albeit with much smaller error bars (see the last column of their Table 1). This is due to the fact that they only used the $\Delta J$ and $\Delta K_{\mathrm{S}}$ magnitude differences, thus discarding the $H$-band measurement, which results in a smaller uncertainty on the spectral type of the contaminating star. As we see no reason to discard any of the three measurements, we chose to adopt the safer procedure outlined above, giving equal weight to the three color indices.

\begin{table}
\centering
\vspace{-0.1cm}
\begin{tabular}{cc}
\hline
Filter & $F_{\mathrm{cont}}/F_{\mathrm{W103}}$\\
\hline
$g'$ & 0.0275 $\pm$ 0.0243 \\
$r'$ & 0.0419 $\pm$ 0.0322 \\
$R$ & 0.0447 $\pm$ 0.0318 \\
$BB$ & 0.0552 $\pm$ 0.0255 \\
$i'$ & 0.0586 $\pm$ 0.0250 \\
$I$ & 0.0633 $\pm$ 0.0236 \\
$z'$ & 0.0800 $\pm$ 0.0145 \\
$K_\mathrm{S}$ & 0.1637 $\pm$ 0.0029 \\
\hline
\end{tabular}
\vspace{-0.1cm}
\caption{Contaminant-to-target flux ratios in the observed passbands.}
\label{dilution}
\end{table}

\vspace{0.2cm}
\subsection{Global data analysis}
\vspace{0.2cm}

\label{analysis_method}

To obtain the strongest constraints on the system parameters, we performed a global Bayesian analysis of the whole dataset (41 eclipse light curves and 23 RVs). We used for this purpose the most recent version of the adaptive Markov Chain Monte-Carlo (MCMC) code described in \citet[and references therein]{gillonmcmc}, that derives the posterior probability distribution functions of the global model parameters, basing on stochastic simulations. Each UT time of mid-exposure was converted to the $\mathrm{BJD}_{\mathrm{TDB}}$ time-scale (\citealt{eastman}). To model the photometry, we used the eclipse model of \cite{mandelagol} multiplied by a different baseline model for each light curve (see below), while the RVs were modeled using a Keplerian orbit (e.g. \citealt{murraycorreia}) combined to a systemic velocity. A quadratic limb-darkening law was assumed for the transits.\\
\indent
The photometric baseline models, different for each light curve, allowed to account for photometric variations not related to the eclipses but rather to external astrophysical, instrumental, or environmental effects. They consisted of different polynomials with respect to, e.g., time, airmass, PSF full-width at half maximum, background, stellar position on the detector, or any combination of these parameters. For each light curve, the optimal baseline function (see Table \ref{tablelog}) was selected by way of minimizing the Bayesian Information Criterion (BIC, \citealt{schwarz}). For the TRAPPIST light curves, a normalization offset was also part of the baseline model to represent the effect of the meridian flip; that is, the $180^{\circ}$ rotation that the German equatorial mount telescope has to undergo when the meridian is reached. This movement results in different positions of the stellar images on the detector before and after the flip, and the normalization offset allows to account for a possible consecutive jump in the differential photometry at the time of the flip.

\indent
The jump parameters in our MCMC analysis (i.e. the parameters that are randomly perturbed at each step of the MCMC) were:
\begin{itemize}
\item the transit depth in the $R$-band \hbox{$\mathrm{d}F_{R}$ = $(R_{\mathrm{p,}R}/R_{\star})^{2}$}, where $R_{\mathrm{p,}R}$ is the planetary radius in the $R$-band and $R_{\mathrm{\star}}$ is the stellar radius;
\item the transit depth differences in the other wavelength bands $\mathrm{dd}F_{j}$ = $\mathrm{d}F_{j} - \mathrm{d}F_{R}$ (where \hbox{$j$ = $g'$,} $r'$, $BB$, $i'$, $I$, $z'$);
\item the occultation depths in the $z'$ and $K_{\mathrm{S}}$-bands, noted $\mathrm{d}F_{\mathrm{occ,}z'}$ and $\mathrm{d}F_{\mathrm{occ,}K_{\mathrm{S}}}$, respectively; 
\item the transit impact parameter in the case of a circular orbit \hbox{$b'$ = $a$ $\mathrm{cos}$ $i_{\mathrm{p}}/R_{\star}$}, where $a$ is the orbital semi-major axis and $i_{\mathrm{p}}$ is the orbital inclination;
\item the transit width (from $1^{\mathrm{st}}$ to $4^{\mathrm{th}}$ contact) $W$;
\item the time of mid-transit $T_{0}$;
\item the orbital period $P$;
\item the stellar effective temperature $T_{\mathrm{eff}}$ and metallicity [Fe/H];
\item the parameter \hbox{$K_{2} = K \sqrt{1-e^{2}}\:\:P^{1/3}$}, where $K$ is the RV orbital semi-amplitude and $e$ is the orbital eccentricity;
\item the two parameters \hbox{$\sqrt{e}$ cos $\omega$} and \hbox{$\sqrt{e}$ sin $\omega$}, where $\omega$ is the argument of the periastron;
\item the linear combinations of the quadratic limb-darkening coefficients ($u_{1,j}$, $u_{2,j}$) in each wavelength band, \hbox{$c_{1,j} = 2 u_{1,j} + u_{2,j}$} and \hbox{$c_{2,j}=u_{1,j} - 2 u_{2,j}$} (where \hbox{$j$ = $g'$,} $r'$, $R$, $BB$, $i'$, $I$, $z'$).
\end{itemize}
The baseline model parameters were not jump parameters; they were determined by linear least-squares minimization from the residuals at each step of the MCMC, thanks to their linear nature
in the baseline functions. For this purpose, a Singular Value Decomposition (SVD) method was used (\citealt{Press1992}). This approach allows to increase significantly the efficiency of the MCMC implementation by reducing the number of jump parameters and thus the dimension of the space to probe.\\
\indent
Normal prior probability distribution functions were assumed for $T_{\mathrm{eff}}$, [Fe/H], $u_{1,j}$, and $u_{2,j}$. For $T_{\mathrm{eff}}$ and [Fe/H], the priors were based on the values reported in \citetalias{gillon14}, with expectations and standard deviations corresponding to the quoted measurements and errors, respectively. As for the normal priors imposed on $u_{1,j}$ and $u_{2,j}$, their parameters were interpolated from the theoretical tables of \cite{claret}. All these normal prior distributions are presented in Table \ref{priors}. Uniform prior distributions were assumed for the other jump parameters. At each MCMC step, a value for $F_{\mathrm{cont}}/F_{\mathrm{W103}}$ for each wavelength band was drawn from the normal distribution having as expectation and standard deviation the value and error given in Table \ref{dilution} for this band, respectively. The dilution correction factor for each band was computed as 1+$F_{\mathrm{cont}}/F_{\mathrm{W103}}$.

\begin{table}
\centering
\begin{tabular}{cc}
\hline
Parameter & Prior\\
\hline
$T_{\mathrm{eff}}$ & $\mathcal{N}(6110,160^{2})$ K \\
$[\mathrm{Fe/H}]$ & $\mathcal{N}(0.06, 0.13^{2})$ dex \\
$u_{\mathrm{1,}g'}$ & $\mathcal{N}(0.502, 0.032^{2})$ \\
$u_{\mathrm{2,}g'}$ & $\mathcal{N}(0.253, 0.020^{2})$ \\
$u_{\mathrm{1,}r'}$ & $\mathcal{N}(0.337, 0.024^{2})$ \\
$u_{\mathrm{2,}r'}$ & $\mathcal{N}(0.305, 0.010^{2})$ \\
$u_{\mathrm{1,}R}$ & $\mathcal{N}(0.316, 0.023^{2})$ \\
$u_{\mathrm{2,}R}$ & $\mathcal{N}(0.304, 0.010^{2})$ \\
$u_{\mathrm{1,}BB}$ & $\mathcal{N}(0.316, 0.046^{2})$ \\
$u_{\mathrm{2,}BB}$ & $\mathcal{N}(0.304, 0.020^{2})$ \\
$u_{\mathrm{1,}i'}$ & $\mathcal{N}(0.260, 0.020^{2})$ \\
$u_{\mathrm{2,}i'}$ & $\mathcal{N}(0.298, 0.008^{2})$ \\
$u_{\mathrm{1,}I}$ & $\mathcal{N}(0.242, 0.020^{2})$ \\
$u_{\mathrm{2,}I}$ & $\mathcal{N}(0.296, 0.008^{2})$ \\
$u_{\mathrm{1,}z'}$ & $\mathcal{N}(0.207, 0.018^{2})$ \\
$u_{\mathrm{2,}z'}$ & $\mathcal{N}(0.290, 0.007^{2})$ \\
\hline
\end{tabular}
\caption{Prior probability distribution functions assumed in our MCMC analysis. $\mathcal{N}(\mu,\sigma^{2})$ represents a normal distribution with an expectation $\mu$ and a variance $\sigma^{2}$.}
\label{priors}
\vspace{-0.2cm}
\end{table}

\indent
The physical parameters of the system were deduced from the jump parameters at each step of the MCMC, so that their posterior probability distribution functions could also be constructed. At each MCMC step, a value for the stellar mean density $\rho_{\star}$ was first derived from the Kepler's third law and the jump parameters $\mathrm{d}F_{R}$, $b'$, $W$, $P$, \hbox{$\sqrt{e}$ cos $\omega$} and \hbox{$\sqrt{e}$ sin $\omega$} (see e.g. \citealt{seager} and \citealt{winn}). This $\rho_{\star}$ and values for $T_{\mathrm{eff}}$ and $[$Fe/H$]$ drawn from their normal prior distributions were used to determine a value for the stellar mass $M_{\star}$ through an empirical law \hbox{$M_{\star}$($\rho_{\star}$, $T_{\mathrm{eff}}$, $[$Fe/H$]$)} (\citealt{enoch}, \citealt{gillon3}) that is calibrated using the set of well-constrained detached eclipsing binary (EB) systems presented by \cite{southworth}. Here, we chose to reduce this set to the 116 stars with a mass between 0.7 and 1.7 $M_{\odot}$, the goal of this selection being to benefit from our preliminary knowledge of the mass of WASP-103 (\citetalias{southworth15} give $M_{\star}$ = 1.204 $\pm$ 0.089 $M_{\odot}$) to improve the determination of the system parameters. In order to propagate correctly the error on the empirical law, the parameters of the selected subset of calibration stars were normally perturbed within their observational error bars and the coefficients of the law were redetermined at each MCMC step. We furthermore took into account the inability of the empirical law to perfectly reproduce the distribution of the stellar masses by determining at each step of the MCMC the quadratic difference between the RMS of the residuals of the modeling of the EB masses by the empirical law and the mean mass error for the EB sample. At each MCMC step, a new value for $M_{\star}$ was drawn from a normal distribution having as expectation and standard deviation the mass originally determined by the empirical law and the quadratic difference mentioned above, respectively. The stellar radius $R_{\star}$ was derived from $M_{\star}$ and $\rho_{\star}$, and the other physical parameters of the system were then deduced from the jump parameters and stellar mass and radius.

\begin{figure}
\centering     
\includegraphics[scale=0.43] {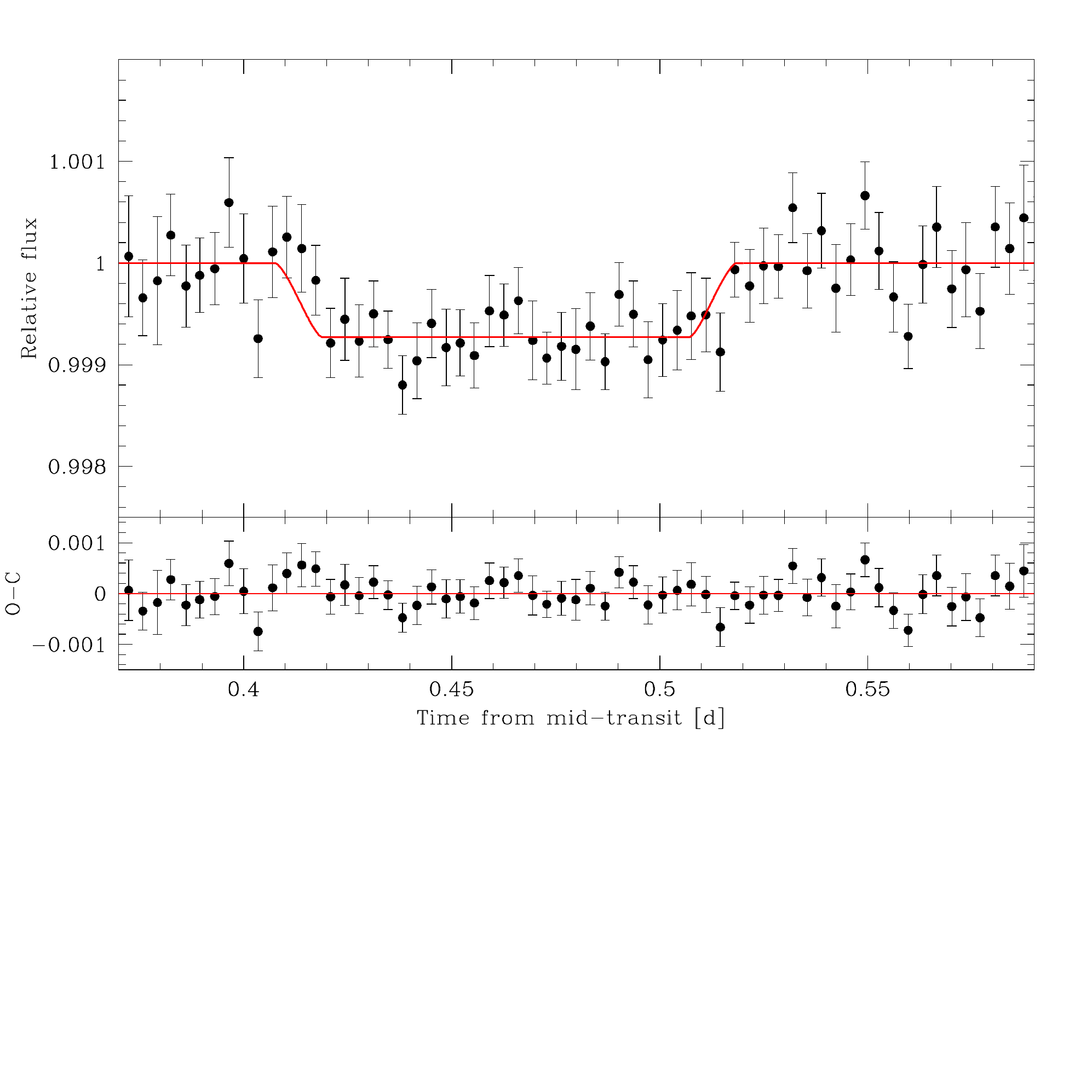} 
\vspace{-3.3cm}
\caption{\textit{Top:} combined occultation photometry obtained in the $z'$-band with TRAPPIST and Euler/EulerCam. The data are period-folded on the best-fit transit ephemeris from our global MCMC analysis (see Section \ref{analysis_method}), corrected for the photometric baseline, and binned in 5min bins (for visual convenience). The best-fit occultation model in the $z'$-band is overplotted in red. The data and model are not corrected for the dilution by the nearby star here. \textit{Bottom:} corresponding residuals. The RMS of the residuals in the shown interval is 305 ppm (5min bins).}     
\label{occ_combz}
%\vspace{-0.5cm}
\end{figure}

\begin{figure*}
\centering     
\includegraphics[scale=0.80] {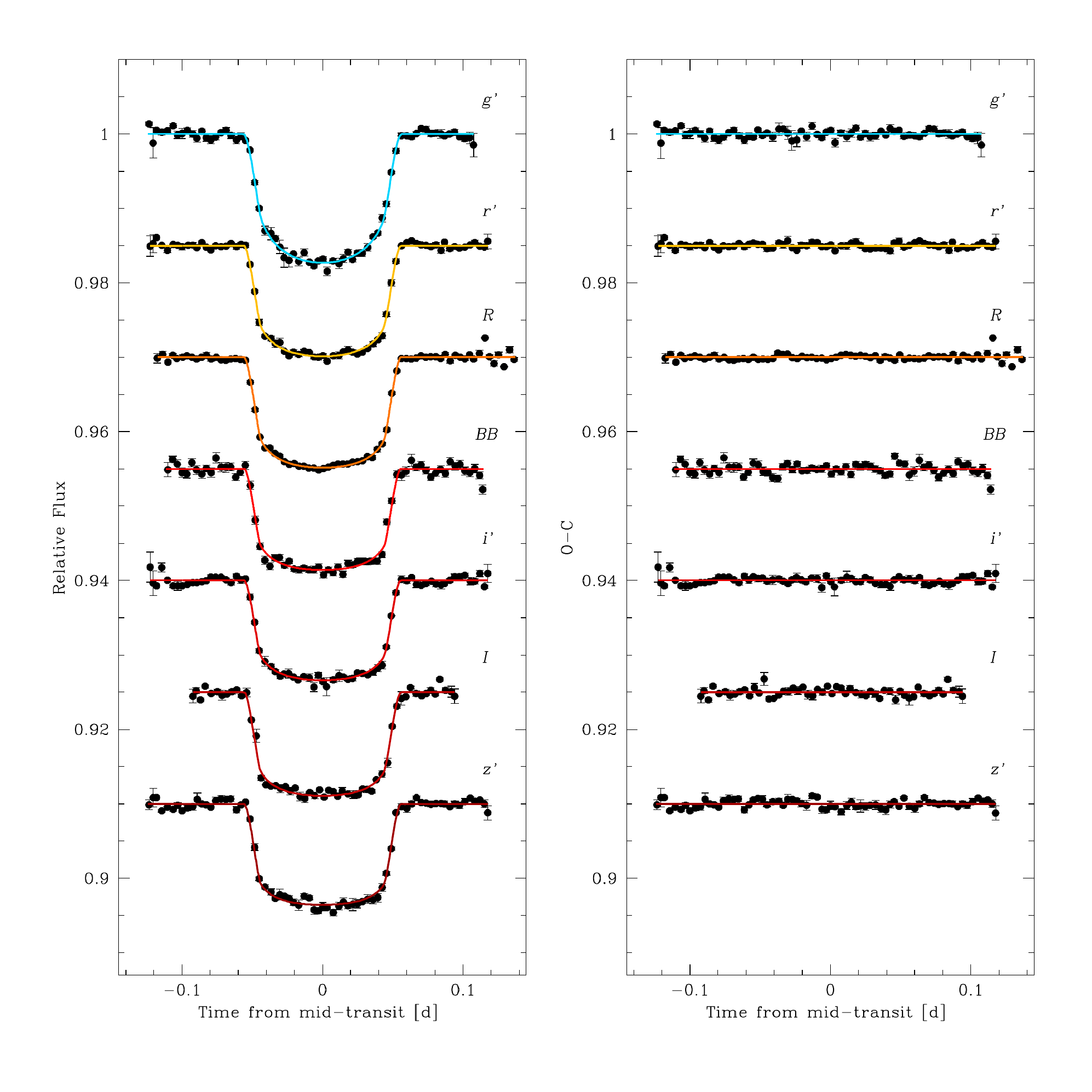} 
\vspace{-1.0cm}
\caption{\textit{Left:} combined transit light curves in each of the observed passbands. The data are period-folded on the best-fit transit ephemeris from our global MCMC analysis (see Section \ref{analysis_method}), corrected for the photometric baseline, and binned in 5min bins (for visual convenience). For each band, the overplotted, solid line is our best-fit transit model. The data and models are not corrected for the dilution by the nearby star here. The light curves are shifted along the \textit{y}-axis for clarity. \textit{Right:} corresponding residuals. The RMS of the residuals in the interval [-0.09,0.09] days are (from top to bottom) 477, 354, 249, 655, 392, 582, and 496 ppm (5min bins).}     
\label{transits_comb}
\vspace{0.3cm}
\end{figure*}

\indent
Although the photometric errors were computed considering scintillation, sky, dark, readout and photon noises, they are known to be often moderately underestimated. A preliminary MCMC analysis, consisting of one chain of \hbox{50 000} steps, was performed to determine the correction factors $CF$ to be applied to the error bars of each photometric time series, as described in \cite{gillonmcmc}. For each light curve, $CF$ is the product of two contributions, $\beta_{w}$ and $\beta_{r}$. On one side, $\beta_{w}$ represents the under- or overestimation of the white noise of each measurement. It is computed as the ratio between the standard deviation of the residuals and the mean photometric error. On the other side, $\beta_{r}$ allows to account for possible correlated noise present in the light curve. It is obtained by comparing the standard deviations of the binned and unbinned residuals for different binning intervals ranging from 5 to 120min, with the largest value being kept as $\beta_{r}$. The values deduced for $\beta_{w}$, $\beta_{r}$, and \hbox{$CF$ = $\beta_{w} \times \beta_{r}$} for each light curve are given in Table \ref{tablelog}. Similarly, this preliminary analysis allowed us to assess the need to rescale the RV error bars, but it was unnecessary here (the best-fit RV model already gives a reduced $\chi^{2}$ = 1.0).\\
\indent
With the corrected photometric error bars, two analyses were then performed: one assuming a circular orbit \hbox{($e$ = 0)} and one with a free eccentricity. Each analysis consisted of three chains of 100 000 steps, whose convergence was checked using the statistical test of \cite{GR}. The first 20\% of each chain was considered as its burn-in phase and discarded. A model comparison based on the Bayes factor, as estimated from the BIC, strongly favored the circular model (Bayes factor of 4915 in its favor) over the eccentric one. We thus adopted the circular orbit as our nominal solution. The corresponding derived system parameters and 1-$\sigma$ error bars are presented in Table \ref{results}. The best-fit eclipse models are shown in Figs. \ref{lcs_Ks} (third panel, occultation model in the $K_{\mathrm{S}}$-band), \ref{occ_combz} (occultation model in the $z'$-band), and \ref{transits_comb} (transit models in each of the observed passbands), while the best-fit RV model is displayed in \hbox{Fig. \ref{rvs}}.

\begin{table*}
\centering
\caption{\textbf{System parameters:} median values and 1-$\sigma$ limits of the posterior probability distribution functions derived from our global MCMC analysis.}
\begin{tabular}{lcl}
\hline
\textbf{Parameters} & \textbf{Values} & \textbf{Units} \\
\hline
\textit{Stellar parameters} & & \\
\hline
Effective temperature $T_{\mathrm{eff}}$ & 6110 $\pm$ 160 & K \\
Metallicity [Fe/H] & 0.06 $\pm$ 0.13 & dex \\
Surface gravity log $g_{\star}$ & $4.219_{-0.014}^{+0.013}$ & cgs \\
Mean density $\rho_{\star}$ & $0.427_{-0.006}^{+0.004}$ & $\rho_{\odot}$ \\
Mass $M_{\star}$ & 1.21 $\pm$ 0.11 & $M_{\odot}$ \\
Radius $R_{\star}$ & 1.416 $\pm$ 0.043 & $R_{\odot}$ \\
\hline
\textit{Planet parameters} & & \\
\hline
Transit depth (in the $R$-band) $\mathrm{d}F_{R}$ = $(R_{\mathrm{p,}R}/R_{\star})^{2}$ & $1.323_{-0.031}^{+0.046}$ & $\%$ \\
Transit impact parameter $b' = a\:\mathrm{cos}\:i_{\mathrm{p}}/R_{\star}$ & $0.06_{-0.05}^{+0.06}$ & $R_{\star}$ \\
Transit width $W$ & 0.1090 $\pm$ 0.0003 & d \\
Time of inferior conjunction $T_{0}$ & 2 456 836.296427 $\pm$ 0.000063 & $\mathrm{BJD_{TDB}}$ \\
Orbital period $P$ & 0.92554517 $\pm$ 0.00000058 & d \\
RV semi-amplitude $K$ & 270 $\pm$ 14 & $\mathrm{m\:s^{-1}}$ \\
Scaled semi-major axis $a/R_{\star}$ & $3.010_{-0.013}^{+0.008}$ & -- \\
Orbital semi-major axis $a$ & $0.01979_{-0.00061}^{+0.00057}$ & AU \\
Orbital inclination $i_{\mathrm{p}}$ & $88.8_{-1.1}^{+0.8}$ & deg \\
Equilibrium temperature$^{a}$ $T_{\mathrm{eq}}$ & 2484 $\pm$ 67 & K \\
Surface gravity log $g_{\mathrm{p}}$ & $3.171_{-0.024}^{+0.027}$ & cgs \\
Mean density $\rho_{\mathrm{p}}$ & $0.353_{-0.024}^{+0.028}$ & $\rho_{\mathrm{Jup}}$ \\
Mass $M_{\mathrm{p}}$ & 1.51 $\pm$ 0.11 & $M_{\mathrm{Jup}}$ \\
Radius (in the $R$-band) $R_{\mathrm{p,}R}$ & $1.623_{-0.053}^{+0.051}$ & $R_{\mathrm{Jup}}$ \\
Roche limit$^{b}$ $a_{\mathrm{R}}$ & $0.01760_{-0.00082}^{+0.00079}$ & AU \\
$a/a_{\mathrm{R}}$ & $1.124_{-0.026}^{+0.029}$ & \\
\hline
\textit{Planet parameters corrected for asphericity (Section \ref{sphere})} & & \\
\hline
Radius (in the $R$-band) $R_{\mathrm{p,}R}$ & 1.681 $\pm$ 0.063 & $R_{\mathrm{Jup}}$ \\
Mean density $\rho_{\mathrm{p}}$ & 0.318 $\pm$ 0.035 & $\rho_{\mathrm{Jup}}$ \\
\hline
\textit{Planet/star radius ratio $R_{\mathrm{p}}/R_{\star}$ (transmission spectrum)} & & \\
\hline
$R_{\mathrm{p,}g'}/R_{\star}$ (0.48 $\mu$m) & $0.1180_{-0.0016}^{+0.0018}$ & -- \\
$R_{\mathrm{p,}r'}/R_{\star}$ (0.62 $\mu$m)  & $0.1155_{-0.0018}^{+0.0016}$ & -- \\
$R_{\mathrm{p,}R}/R_{\star}$ (0.66 $\mu$m)  & $0.1150_{-0.0014}^{+0.0020}$ & -- \\
$R_{\mathrm{p,}BB}/R_{\star}$ (0.70 $\mu$m)$^{c}$ & $0.1109_{-0.0018}^{+0.0024}$ & -- \\
$R_{\mathrm{p,}i'}/R_{\star}$ (0.76 $\mu$m) & $0.1116_{-0.0019}^{+0.0023}$ & -- \\
$R_{\mathrm{p,}I}/R_{\star}$ (0.82 $\mu$m) & 0.1133 $\pm$ 0.0013 & -- \\
$R_{\mathrm{p,}z'}/R_{\star}$ (0.90 $\mu$m) & $0.1143_{-0.0012}^{+0.0013}$ & -- \\
\hline
\textit{Occultation depths $\mathrm{d}F_{\mathrm{occ}}$ (emission spectrum)} & & \\
\hline
$\mathrm{d}F_{\mathrm{occ,}z'}$ (0.90 $\mu$m) & 699 $\pm$ 110 & ppm \\
$\mathrm{d}F_{\mathrm{occ,}K_{\mathrm{S}}}$ (2.15 $\mu$m) & $3567_{-350}^{+400}$ & ppm \\
\hline
\end{tabular}
\begin{tablenotes}
\item[] 
\textbf{Notes.} $ ^{a}$Assuming a null Bond albedo.
\item[] 
$ ^{b}$Using \hbox{$a_{\mathrm{R}}$ = 2.46 $R_{\mathrm{p}}(M_{\star}/M_{\mathrm{p}})^{1/3}$} (\citealt{chandr}).
\item[] 
$ ^{c}$TRAPPIST blue-blocking filter.
\end{tablenotes}
\label{results}
\end{table*}

\section{Discussion}
\label{discuss}

\vspace{-0.2cm}

\subsection{System physical parameters and correction for asphericity}
\label{sphere}

As expected, we find a slightly larger radius ($1.623_{-0.053}^{+0.051}$ $R_{\mathrm{Jup}}$) and a slightly lower mean density ($0.353_{-0.024}^{+0.028}$ $\rho_{\mathrm{Jup}}$) for the planet than those reported by \citetalias{gillon14} ($1.528_{-0.047}^{+0.073}$ $R_{\mathrm{Jup}}$, $0.415_{-0.053}^{+0.046}$ $\rho_{\mathrm{Jup}}$) and \citetalias{southworth15} ($1.554\pm0.044$ $R_{\mathrm{Jup}}$, $0.367\pm0.027$ $\rho_{\mathrm{Jup}}$), who did not take into account the contamination from the nearby star, unknown at that time. Our values for these two parameters agree well with those found by \citetalias{southworth16} ($1.596_{-0.054}^{+0.044}$ $R_{\mathrm{Jup}}$, $0.339\pm0.023$ $\rho_{\mathrm{Jup}}$). The other physical parameters of the system are in very good agreement with those reported by \citetalias{gillon14}, \citetalias{southworth15}, and \citetalias{southworth16}.\\
\indent
With an orbital semi-major axis only $\sim$1.12 times larger than its Roche limit, WASP-103\,b is expected to be significantly deformed by tides (e.g. \citealt{budaj11}). We calculated values for the planetary radius and mean density corrected for asphericity using the same method as \citetalias{southworth15} (also applied to the case of WASP-121\,b by \citealt{delrez121}). In brief, we used the Roche model of \cite{budaj11} to compute the Roche shape of the planet which would have the same cross-section during transit as the one we inferred from our observations assuming a spherical planet (eclipse model of \citealt{mandelagol}, see Section \ref{analysis_method}). The main inputs of the model were the orbital semi-major axis ($a=4.26_{-0.13}^{+0.12}$ $R_{\odot}$), the star-to-planet mass ratio ($M_{\star}/M_{\mathrm{p}}=840\pm137$), and the planetary radius obtained assuming a spherical shape ($R_{\mathrm{p,}R}=1.623_{-0.053}^{+0.051}$ $R_{\mathrm{Jup}}$). We found a corrected value for the planetary radius of $1.681\pm0.063$ $R_{\mathrm{Jup}}$ (radius of the sphere that would have the same volume as the Roche surface of the planet) and a corresponding revised mean density of $0.318\pm0.035$ $\rho_{\mathrm{Jup}}$. These corrected values are also included in Table \ref{results}.

\subsection{Atmospheric properties of WASP-103\,b}

We use our data to constrain the atmospheric properties of WASP-103\,b. For this purpose, we modeled the atmospheric spectra of WASP-103\,b, both in thermal emission and in transmission, using an exoplanetary atmospheric modeling and retrieval method based on \cite{madhu2009} (also see \citealt{madhu2011}, \citealt{madhu2012}). Here, we briefly summarize the method. The model consists of a 1-D radiative transfer solver which computes the observed spectrum of an exoplanetary atmosphere for a given geometry, in transit or at secondary eclipse, assuming plane parallel geometry. The temperature profile and chemical composition of the atmosphere are free parameters of the model, with 6 parameters for the $P-T$ profile and \hbox{4-6} free parameters for the chemical species; one parameter for each relevant atom/molecule. We include the major opacity sources expected in hot hydrogen-dominated atmospheres, namely H$_2$O, CO, CH$_4$, CO$_2$, C$_2$H$_2$, HCN, TiO, VO, and collision-induced absorption (CIA) due to H$_{2}-$H$_2$, as described in \cite{madhu2012}. The model assumes hydrostatic equilibrium and local thermodynamic equilibrium (LTE) and while computing thermal emission spectra ensures global energy balance with the incident radiation. The parametric temperature structure and molecular abundances in the model allow exploration of a wide range of temperature profiles and chemical compositions in search of the best-fit models to fit the data.

\subsubsection{Emission spectrum}

We clearly detect the thermal emission from the dayside of the planet in both the $z'$ (0.9 $\mu$m) and $K_{\mathrm{S}}$ (2.1 $\mu$m) bands, the measured occultation depths being 699$\pm$110 ppm (6.4-$\sigma$ detection) and \hbox{$3567_{-350}^{+400}$ ppm} (10.2-$\sigma$ detection), respectively. These two measurements suggest a peculiar feature. The occultation depth in the $K_{\mathrm{S}}$-band corresponds to a brightness temperature ($T_{\mathrm{B}}$) of $3171_{-130}^{+144}$ K which is marginally higher ($\sim$1.7-$\sigma$) than the $T_{\mathrm{B}}$ obtained in the \hbox{$z'$-band} of $2914_{-87}^{+80}$ K. Generally, for hot Jupiters, the $T_{\mathrm{B}}$ in the $z'$-band is found to be consistent with or higher than that in the $K_{\mathrm{S}}$-band (see e.g. \citealt{anderson2013} and \citealt{lendl2013} for \hbox{WASP-19\,b}, \citealt{madhu2012} for \hbox{WASP-12\,b}, or \citealt{haynes2015} for WASP-33\,b). This is because the \hbox{$z'$-band} contains strong spectral features due to TiO. On the other hand, the $K_{\mathrm{S}}$-band is relatively devoid of strong spectral features due to TiO and other molecules relevant for hot Jupiters, with the exception of some weak features due to CO. Thus, when TiO is absent all the near-infrared ground-based photometric bands ($z'$, $J$, $H$, and $K_{\mathrm{S}}$) provide windows in atmospheric opacity and probe the temperatures in the deep atmosphere which tends to be isothermal for hot Jupiters. Therefore, the brightness temperatures in all these channels are expected to be similar, as observed for several hot Jupiters (e.g. \citealt{anderson2013}, \citealt{lendl2013}, \citealt{madhu2012}). On the other hand, in hot Jupiters where TiO is present, it can lead to a thermal inversion in the atmosphere (e.g. \citealt{fortney2008}) and cause spectral features in emission. In this case, the TiO features in the $z'$-band lead to a higher $T_{\mathrm{B}}$ in that band compared to that in the $K_{\mathrm{S}}$-band which still has limited opacity (e.g. \citealt{haynes2015}). Therefore, it is rarely the case that the $T_{\mathrm{B}}$ in the $z'$-band is lower than that in the $K_{\mathrm{S}}$-band, assuming a solar composition atmosphere.

\begin{figure}
\centering     
\includegraphics[scale=0.55] {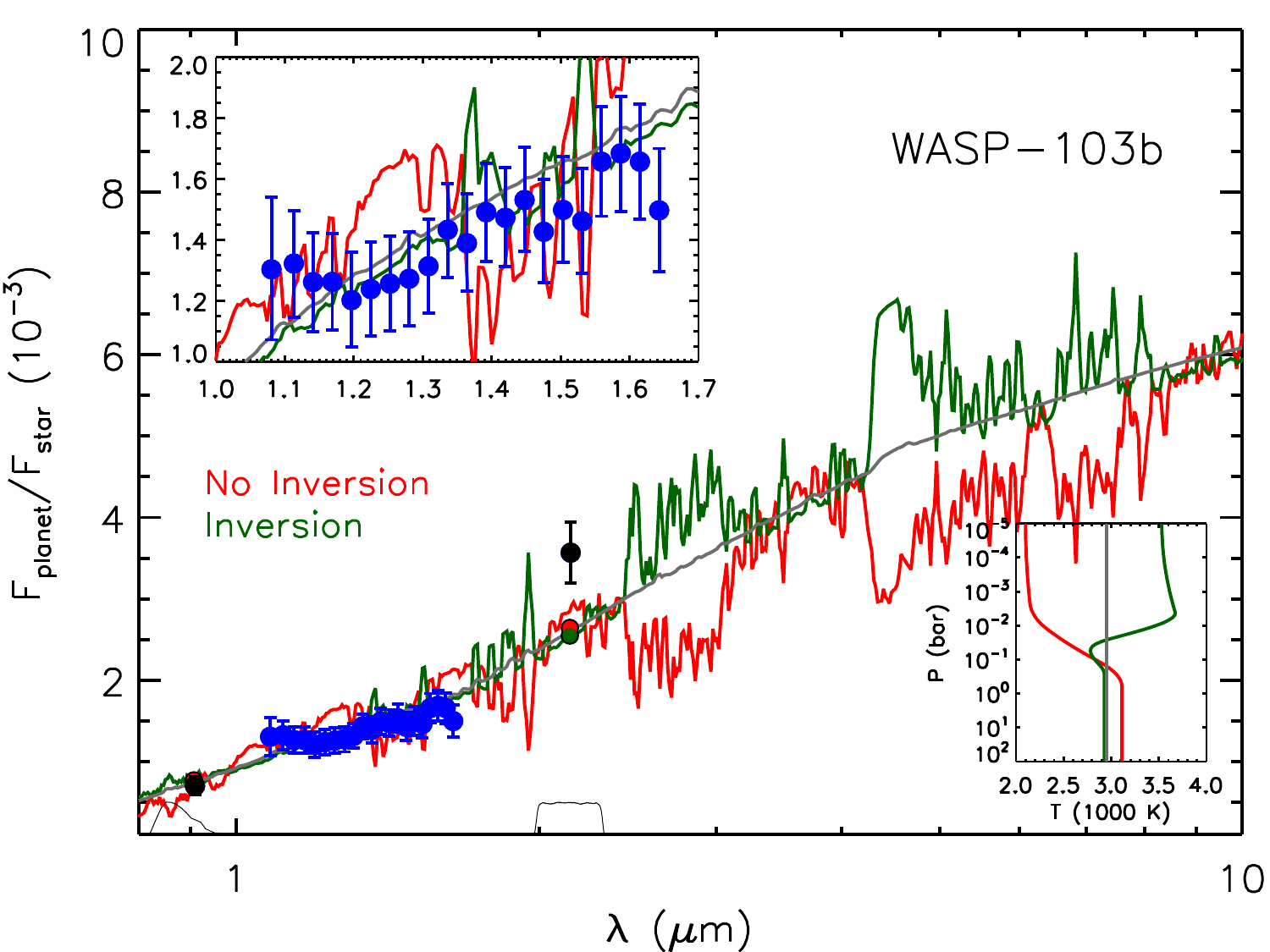} 
\vspace{-0.5cm}
\caption{Observations and model spectra of dayside thermal emission from WASP-103\,b. The black circles with error bars show our $z'$ and $K_{\mathrm{S}}$-band measurements, while the HST/WFC3 data reported by \citetalias{cartier17} are plotted in blue. A zoom on the HST/WFC3 measurements is shown in the top inset. The colored curves show best-fit model spectra corresponding to three model scenarios: a blackbody model with a temperature of 2900 K (grey), a model with a thermal inversion and a solar composition but a low TiO abundance (green), and a 0.5$\times$solar model without a thermal inversion (red). The bottom inset shows the corresponding pressure-temperature profiles for the models. The green and red circles give the band-integrated fluxes of the corresponding models in the observed photometric bands (bottom black curves), for comparison to the data.}     
\label{spectraem}
\end{figure}

\indent
Fig.~\ref{spectraem} shows our occultation measurements along with model emission spectra of WASP-103\,b. The HST/WFC3 measurements recently reported by \citetalias{cartier17} are also shown. We consider three model scenarios in order to explain the data. Firstly, we find that the $z'$-band and WFC3 data are very well explained by a featureless blackbody spectrum, with a temperature of $\sim$2900 K (grey model in Fig.~\ref{spectraem}). A blackbody spectrum is possible if either the atmosphere is isothermal, as shown here, or the metallicity is very low, i.e. providing very low molecular opacity irrespective of the temperature profile. However, the $K_{\mathrm{S}}$-band point is inconsistent with this blackbody model at $\sim$3-$\sigma$. Secondly, a model with a thermal inversion (green model in Fig.~\ref{spectraem}) can fit the current data at nearly the same level as the blackbody. Here, the inversion model has a solar composition atmosphere, but with 0.1$\times$solar TiO, and an inverted $P-T$ profile shown in the inset of Fig.~\ref{spectraem}. The inversion model is able to achieve such a fit because the continuum of this model is at the same temperature as the blackbody model, with only a few strong emission features in the 1.4 $\mu$m H$_2$O band due to a moderately steep $P-T$ profile. Thirdly, a 0.5$\times$solar non-inverted model (red model in Fig.~\ref{spectraem}) provides a slightly worse fit to the data. The non-inverted model has a higher continuum than the blackbody and strong H$_2$O absorption, neither of which is observed in the WFC3 spectrum. In summary, an isothermal temperature profile and/or a low H$_2$O abundance atmosphere provide the best fits to the $z'$-band and WFC3 data. A low H$_2$O abundance is achievable with either a low metallicity or a high C/O ratio (\citealt{madhu2012}, \citealt{moses2013}). Our model inferences are consistent with those of \citetalias{cartier17} who based their results on the WFC3 data alone.\\ 
\indent
On the other hand, as can be seen in Fig.~\ref{spectraem}, our $K_{\mathrm{S}}$-band data point is matched only at 2-3 $\sigma$ by our current models. To reproduce this high $K_{\mathrm{S}}$-band flux, a strong source of opacity in this wavelength band would be needed. Such a scenario would also require a strong thermal inversion, as well as low H$_2$O and TiO abundances. A thermal inversion would be needed to produce a $K_{\mathrm{S}}$-band emission feature from the above opacity source over the continuum blackbody which is required in the WFC3 band. And, the low H$_2$O and TiO abundances would be required to satisfy the lack of features in the WFC3 and $z'$-bands, respectively. However, as pointed out previously, the $K_{\mathrm{S}}$-band is not expected to contain strong spectral features due to any of the molecules relevant for a hot-Jupiter atmosphere, making the above scenario rather unlikely. Further observations in the $K_{\mathrm{S}}$-band are thus required to confirm our measurement, which is based on a single occultation light curve, before any adequate interpretation of the data can be given.\\
\indent
Additional data are necessary to better characterize the dayside atmosphere of WASP-103\,b. Most notably, observations with \textit{Spitzer} can distinguish between our three model scenarios. The \textit{Spitzer}/IRAC photometric bands at 3.6 and \hbox{4.5 $\mu$m} together should be able to provide strong constraints on the CO absorption versus emission features between the non-inverted and inverted models in the 4-5 micron region. The joint constraints on the CO and H$_2$O abundances could then help constrain the C/O ratio of the dayside atmosphere (\citealt{madhu2012}). As mentioned above, we also encourage further observations in the $K_{\mathrm{S}}$-band to confirm our high emission measurement in that band. On a longer time frame, the \textit{James Webb Space Telescope} (JWST) should be able to provide conclusive constraints on the $P-T$ profile and composition of this planetary atmosphere.

%\indent
%On the other hand, our high $K_{\mathrm{S}}$-band data point suggests the possibility of a new absorber and a thermal inversion. The $K_{\mathrm{S}}$-band flux is matched only at 2-3 $\sigma$ by our current models. It is possible that this high $K_{\mathrm{S}}$-band flux is caused by a strong source of opacity which is not accounted for in our present models. Such a scenario would also require a strong thermal inversion, as well as low H$_2$O and TiO abundances. A thermal inversion is needed in such a scenario to produce a $K_{\mathrm{S}}$-band emission feature from the candidate molecule over the continuum blackbody which is required in the WFC3 band. And, the low H$_2$O and TiO abundances are required to satisfy the lack of features in the WFC3 and $z'$-bands, respectively. Given the high equilibrium temperature of the planet \hbox{($T_{\rm eq}$=2484$\pm$67 K)}, the atmosphere is expected to be conducive for gaseous TiO and VO which in turn could cause a thermal inversion, under the assumption of an oxygen-rich hot Jupiter atmosphere in thermochemical equilibrium (e.g. \citealt{fortney2008}). Other prominent molecules in H$_2$-rich atmospheres at this temperature include H$_2$O and CO. Therefore, the low H$_2$O and TiO abundances would strongly indicate either a low metallicity or a high C/O ratio (\citealt{madhu2011w12}, \citealt{madhu2012}) and an inversion caused by a species other than TiO.\\

%\vspace{0.1cm}

\subsubsection{Transmission spectrum}

%\vspace{0.1cm}

%We use our observed transmission spectrum to constrain the atmospheric properties of WASP-103\,b at the day-night terminator region. 

\begin{figure}
\vspace{-0.2cm}
\centering     
\includegraphics[scale=0.43] {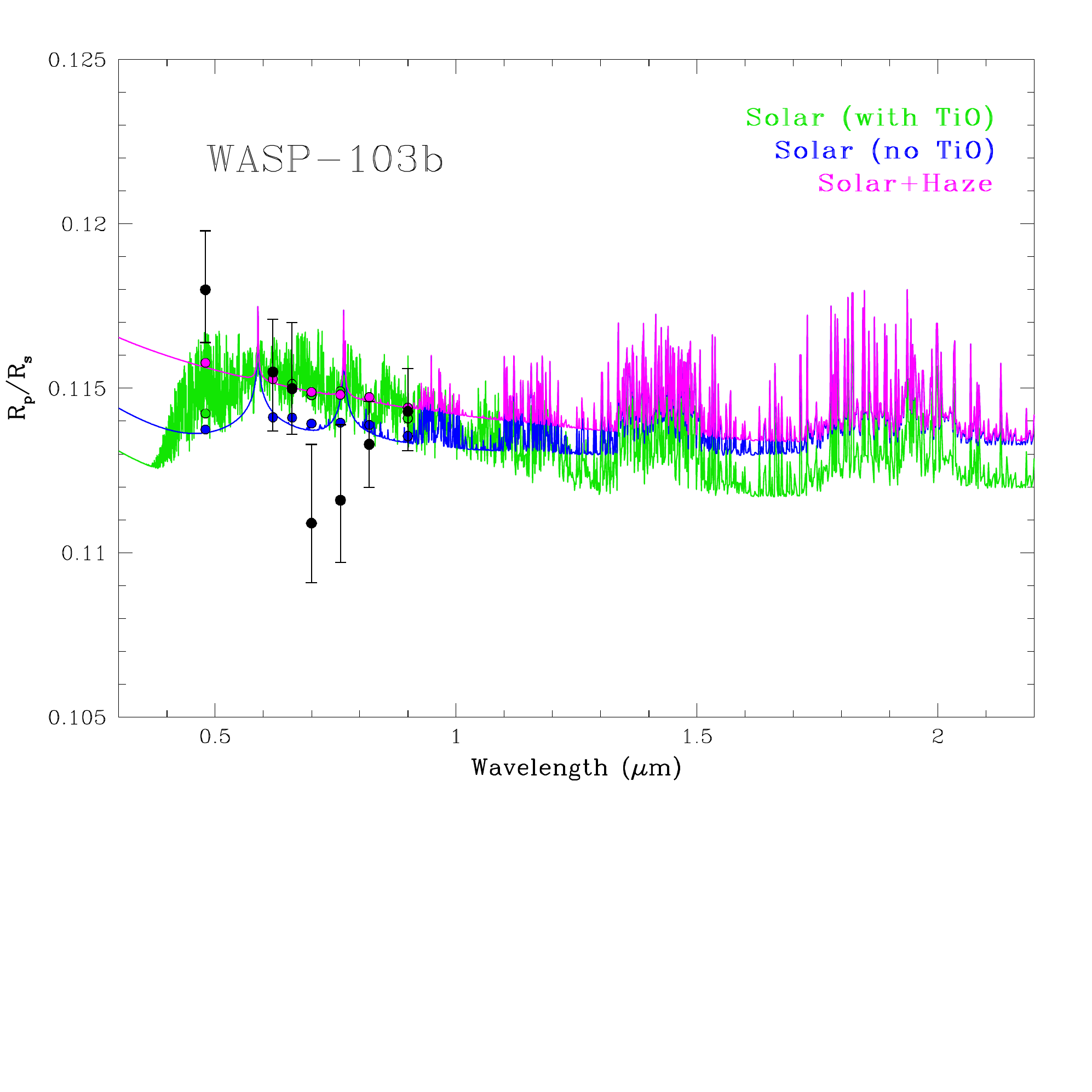} 
\vspace{-2.8cm}
\caption{Broad-band measurements of the planet-to-star radius ratio $(R_{\mathrm{p}}/R_{\star})$ as a function of wavelength compared to model transmission spectra of WASP-103\,b. The various measurements are shown as black circles with error bars. The colored curves show three different plausible models: a fiducial model with solar abundance composition in thermochemical equilibrium (TE) without TiO (blue), a solar abundance model in TE with TiO (green), and a solar abundance model with enhanced scattering due to a possible haze with a scattering index of -3.5 (magenta). The colored circles give the band-integrated fluxes of the corresponding models in the observed photometric bands, for comparison to the data.}     
\label{spectratrans}
\end{figure}

The planet-to-star radius ratios $(R_{\mathrm{p}}/R_{\star})$ obtained for each of the observed passbands are given in Table \ref{results} and shown in Fig. \ref{spectratrans}. We find the same ``V-shape'' pattern as \citetalias{southworth16}, with a minimum effective planetary radius around \hbox{700 nm} and increasing values towards both shorter and longer wavelengths. This pattern is however less significant in our transmission spectrum, due to our more conservative error bars (cf. Section \ref{contam}).

\indent
We compared our broad-band measurements to different model transmission spectra of \hbox{WASP-103\,b}, with the aim to derive constraints on the atmospheric properties of the planet at the day-night terminator region. The possible sources of opacity in the spectral range covered (0.4$-$1.0 $\mu$m) for a typical hot Jupiter in the temperature regime of WASP-103\,b, with an equilibrium temperature of $\sim$2500 K are: (\textit{a}) H$_2$O at the redder wavelengths, (\textit{b}) TiO and VO over the entire range, (\textit{c}) Na and K doublet line opacity peaking at 0.59 $\mu$m and \hbox{0.78 $\mu$m}, respectively, (\textit{d}) Rayleigh scattering due to H$_2$, and (\textit{e}) potential high-temperature refractory hazes, e.g. of Fe particles. Amongst all these sources of opacity, the most prominent sources are TiO and scattering. Given the broad-band photometric nature of the data, we are unable to resolve any particular spectral features. However, firstly, the prominent sources of opacity, e.g. TiO, are significantly broad to influence the data. Secondly, the full visible coverage of the data allows us to investigate a possible blueward slope in the data which, in turn, could constrain the sources of scattering in the atmosphere, including the presence of hazes.\\
\indent
We explored the model parameter space of the opacity sources discussed above in search of models that can explain the data. Fig. \ref{spectratrans} shows three models with different degrees of fit: 
\begin{itemize}
\item a fiducial model with solar abundance composition in thermochemical equilibrium (TE), but with no TiO (blue); 
\item a solar abundance model in TE with TiO (green); 
\item a solar abundance model with enhanced scattering due to a possible haze with a scattering index of -3.5, instead of -4 for H$_{2}$ Rayleigh scattering (magenta). 
\end{itemize}
The corresponding $\chi^{2}$ values are given in Table \ref{chi}. When considering the entire dataset, we find that the model with haze provides a better fit to the data than the two other models. However, we note cautiously that this result relies heavily on the high $R_{\mathrm{p}}/R_{\star}$ measured in the bluest wavelength band in our data, i.e. the $g'$-band. When this data point is not considered, the best fit is obtained by the fiducial solar-composition model without TiO (Table \ref{chi}, last column). It is clear however that none of these models match the data well.\\ 
\indent
During the refereeing process of this paper, a higher resolution optical transmission spectrum of WASP-103\,b obtained with Gemini/GMOS was published by \cite{lendl2017}. These data, covering the wavelength range between 550 and \hbox{960 nm}, do not show any signs of the V-shape pattern found in our measurements. Instead, they show increased absorption in the cores of the Na and K line features, without any other evident trend, thus pointing to a rather clear atmosphere for \hbox{WASP-103\,b} at the pressure levels probed (between 0.01 and 0.1 bar). We have no satisfactory explanation for this discrepancy at this point. Our transmission spectrum is mostly based on the dataset published by \citetalias{southworth15} and re-analysed by \citetalias{southworth16} (17 out of the 25 transit light curves we used). Our independent data analysis gives similar results to theirs, suggesting that the unusual profile of the measured transmission spectrum is intrinsic to the data and not related to the data analysis process.

\begin{table}
\centering
\begin{tabular}{ccc}
\hline
Model & All & Without the $g'$-band\\
 & data & measurement\\
\hline
Solar without TiO & 9.6 & 4.0 \\
Solar with TiO & 10.3 & 6.0 \\
Solar with haze & 7.5 & 5.9 \\
\hline
\end{tabular}
\caption{$\chi^{2}$ values calculated from the data and the various models of the transmission spectrum of WASP-103\,b, considering all data (second column) and omitting the $g'$-band measurement (third column).}
%\vspace{-0.5cm}
\label{chi}
\end{table}

%Our current results pave the way for more detailed studies of WASP-103\,b in the future via transmission spectrophotometry. Future observations with higher spectral resolution and SNR would be able to provide stronger constraints on the planet's atmospheric properties at the day-night terminator region and verify the presence of haze in the planet's atmosphere which is suggested by the current data. The detection of haze in this atmosphere would provide a new data point in support of the presence of very high temperature hazes in hot Jupiters. 

Extending WASP-103\,b's measured transmission spectrum towards near-infrared wavelengths, for example with HST/WFC3, would allow a more detailed characterization of its atmosphere at the day-night terminator region by constraining its H$_2$O abundance. Additional data at shorter wavelengths than the spectral range covered by the GMOS data (i.e. bluewards of 550 nm) would also be useful to definitely assess the presence of a scattering slope possibly related to hazes in \hbox{WASP-103\,b's} transmission spectrum. At the high temperature ($\gtrsim$2000 K) of WASP-103\,b, most of the usual condensate species are in gas phase (see e.g. \citealt{marley2013}), with the exception of few such as Fe and Al$_2$O$_3$. Characterizing the presence of hazes in such atmospheres is important to be able to make robust determinations of chemical abundances from future spectroscopic data using HST and JWST.

\vspace{-0.2cm}

\section{Conclusions}
\label{concl}

In this work, we presented a total of nineteen new eclipse light curves for the ultra-short-period hot Jupiter \hbox{WASP-103\,b}. Sixteen of these light curves were obtained during occultations and three during transits. We also obtained five new RV measurements. We combined these new observations with previously published data and performed a global MCMC analysis of the resulting extensive dataset (41 eclipse light curves and 23 RVs), taking into account the contamination from a faint nearby star.\\
\indent
Using the approach presented in \cite{lendl2013}, that involves combining a large number (here fifteen) of occultation light curves obtained with $\sim$1m-class telescopes to mitigate the effects of correlated noise and progressively extract the occultation signal from the noise, we detected the dayside emission of the planet in the $z'$-band at better than 6-$\sigma$ \hbox{(699$\pm$110 ppm)}. From a single occultation light curve acquired with the CFHT/WIRCam facility, we also detected the planet's dayside emission in the $K_{\mathrm{S}}$-band at better than 10-$\sigma$, the measured occultation depth being \hbox{$3567_{-350}^{+400}$ ppm}. We compared these two measurements, along with recently published HST/WFC3 spectrophotometric data, to model emission spectra of WASP-103\,b with different temperature profiles and chemical compositions. On one hand, we found that the $z'$-band and WFC3 data are best fit by an isothermal atmosphere at a temperature of $\sim$2900 K, or an atmosphere with a low metallicity or a high C/O ratio. On the other hand, we found an unexpectedly high flux in the $K_{\mathrm{S}}$-band when compared to these atmospheric models, which requires confirmation with additional observations before any interpretation can be given.\\
%On the other hand, we found an excess in the $K_{\mathrm{S}}$-band measured flux compared to these models, suggesting the possibility of an emission feature in that band from a new absorber not accounted for in current atmospheric models. Under such a scenario, the atmosphere would thus have a thermal inversion as well as low H$_2$O and TiO abundances (in order to explain the WFC3 and $z'$-band data), which would indicate either a low metallicity or a high C/O ratio and a thermal inversion caused by an atmospheric compound other than TiO.\\
\indent
From our global data analysis, we also derived a broad-band optical transmission spectrum that shows a minimum around 700 nm and increasing values towards both shorter and longer wavelengths. This is in agreement with the results of the study by \citetalias{southworth16}, which was based on a large fraction of the archival transit light curves included in our analysis. The unusual profile of this transmission spectrum is poorly matched by theoretical spectra and is not confirmed by more recent observations at higher spectral resolution reported by \cite{lendl2017}.\\
\indent
Future observations with existing (HST, \textit{Spitzer}) or planned (JWST) facilities, both in emission and transmission, should be able to provide better constraints on the $P-T$ profile and chemical composition of WASP-103\,b's atmosphere. In particular, we encourage further observations in the $K_{\mathrm{S}}$-band to confirm our high emission measurement in that band. Improving our understanding of this planetary system also requires a more precise characterization of the faint nearby star, by obtaining additional adaptive-optics observations with a large-aperture telescope.

%We also used the broad-band optical transmission spectrum derived from our global analysis to gain first insights into the planet's atmospheric properties at the day-night terminator region. We found that the data are better explained by an atmosphere model with enhanced scattering due to haze. However, this result hinges critically on a single data point, the $g'$-band measurement, and should thus be considered tentative.\\

%We also compared the broad-band optical transmission spectrum derived from our global analysis to various model transmission spectra of WASP-103\,b. 

%We found our measurements to be a poor match to these models, with a minimum effective planetary radius around 700 nm and increasing values towards both shorter and longer wavelengths.
%We found a minimum of opacity around 700 nm and increasing values towards both shorter and longer wavelengths, in agreement with the results previously reported by \citetalias{southworth16} which are based on a large subset of the transit light curves used in our work.

%This spectrum is in agreement with the one obtained by \citetalias{southworth16} based on a large fraction of the transit light curves included in our analysis.
%from which most of the transit light curves used in our analysis were taken

%from which a large fraction of the transit light curves used in our work /included in our analysis come from

%are not confirmed by the more recent higher resolution spectrum published by Lendl
%there is no evidence for this pattern in the higher resolution transmission spectrum recently published by

\section*{Acknowledgements}
The research leading to these results has received funding from the European Community's Seventh Framework Programme (FP7/2013-2016) under grant agreement number 312430 (OPTICON). The Canada-France-Hawaii Telescope (CFHT) is operated by the National Research Council of Canada, the Institut National des Sciences de l'Univers of the Centre National de la Recherche Scientifique of France, and the University of Hawaii. The authors thank the CFHT staff, especially Pascal Fouqu\'e, for scheduling, helping prepare, and conducting the CFHT/WIRCam observations used in this work. TRAPPIST is a project funded by the Belgian Fund for Scientific Research (Fonds National de la Recherche Scientifique, F.R.S.-FNRS) under grant FRFC 2.5.594.09.F, with the participation of the Swiss National Science Fundation (SNF). The Swiss {\it Euler} Telescope is operated by the University of Geneva, and is funded by the Swiss National Science Foundation. L. Delrez acknowledges support from the Gruber Foundation Fellowship. M. Gillon and \hbox{E. Jehin} are F.R.S.-FNRS Research Associates. The authors thank the anonymous reviewers for their valuable suggestions.

%%%%%%%%%%%%%%%%%%%%%%%%%%%%%%%%%%%%%%%%%%%%%%%%%%



%\newpage

\bibliographystyle{mnras}
\bibliography{mnras_template} % if your bibtex file is called example.bib

%%%%%%%%%%%%%%%%%%%%%%%%%%%%%%%%%%%%%%%%%%%%%%%%%%

%%%%%%%%%%%%%%%%% APPENDICES %%%%%%%%%%%%%%%%%%%%%

\appendix


\section{Individual eclipse light curves}

\begin{figure*}
\centering     
\includegraphics[scale=0.90] {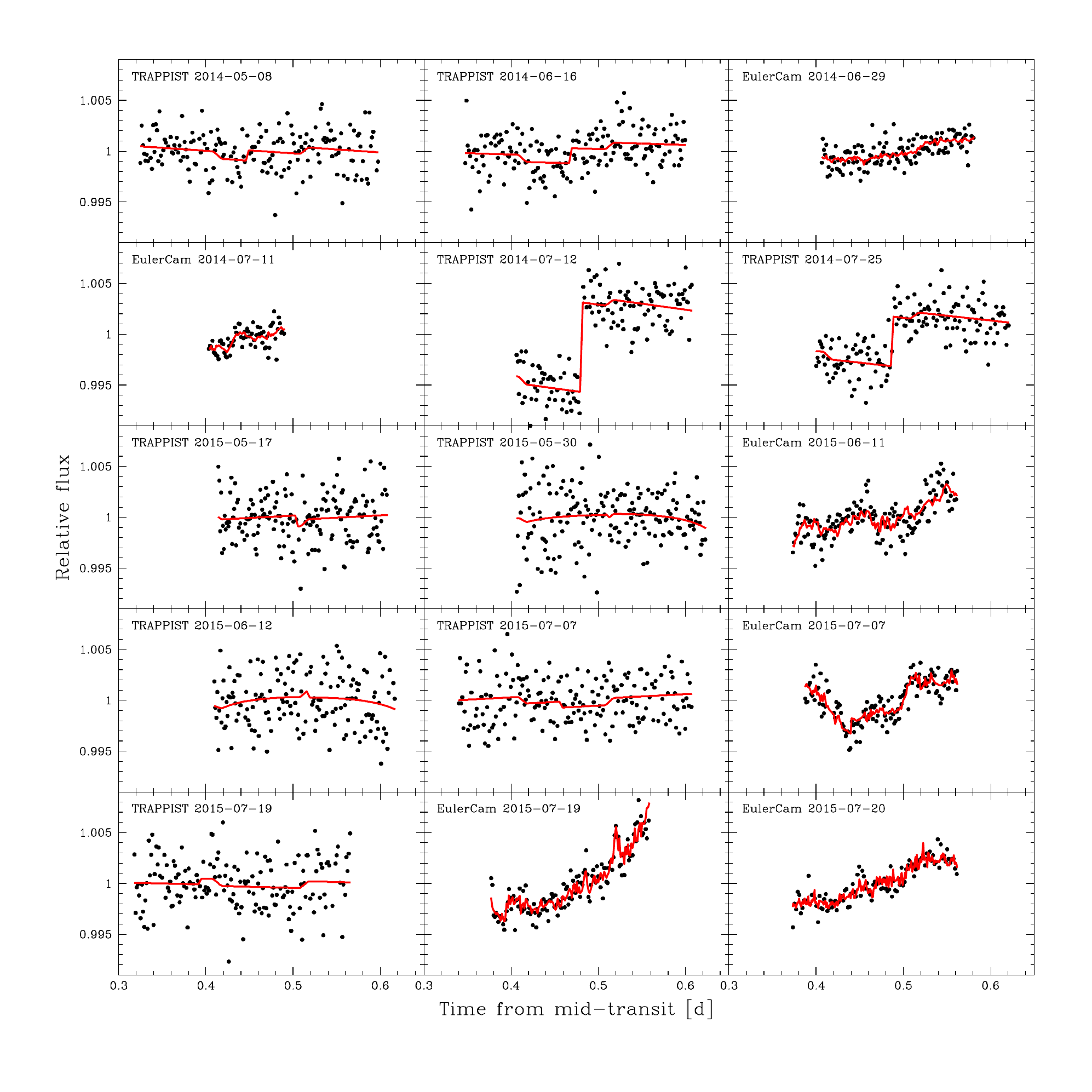} 
\vspace{-1.5cm}
\caption{All raw occultation light curves obtained in the $z'$-band with TRAPPIST and Euler/EulerCam. The data are period-folded on the best-fit transit ephemeris from our global MCMC analysis (see Section \ref{analysis_method}) and binned in 2min bins. For each light curve, the best-fit full model (photometric baseline $\times$ occultation in the $z'$-band) is overplotted in red. The data and models are not corrected for the dilution by the nearby star here.}     
\label{phot_occz}
\end{figure*}

\begin{figure*}
\centering     
\includegraphics[scale=0.90] {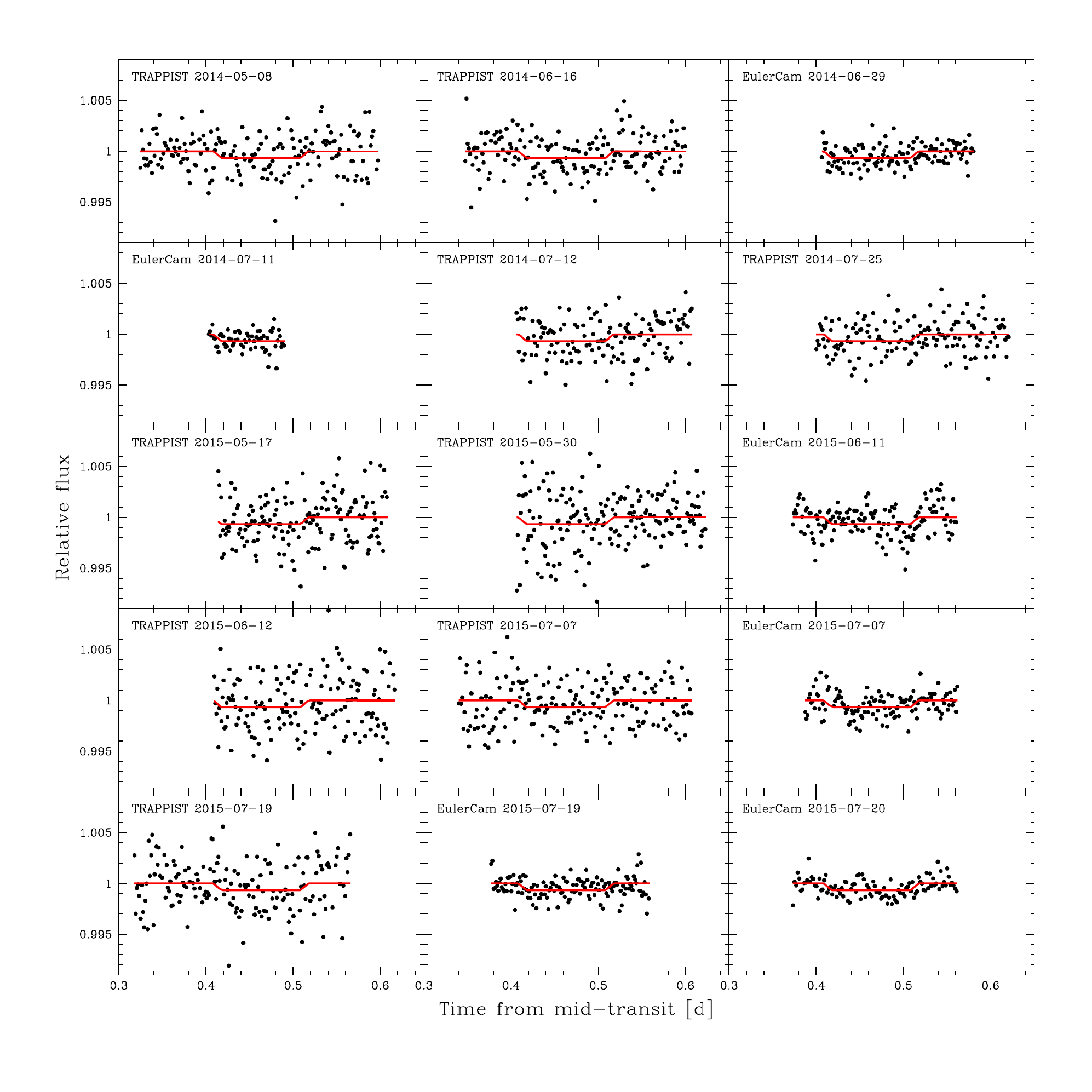} 
\vspace{-1.5cm}
\caption{Same light curves as in Fig. \ref{phot_occz}, but divided by their best-fit baseline models (different for each light curve). The best-fit occultation model in the $z'$-band is overplotted in red. The data and model are not corrected for the dilution by the nearby star here.}     
\label{phoc_occz}
\end{figure*}

\begin{figure*}
\centering     
\includegraphics[scale=0.90] {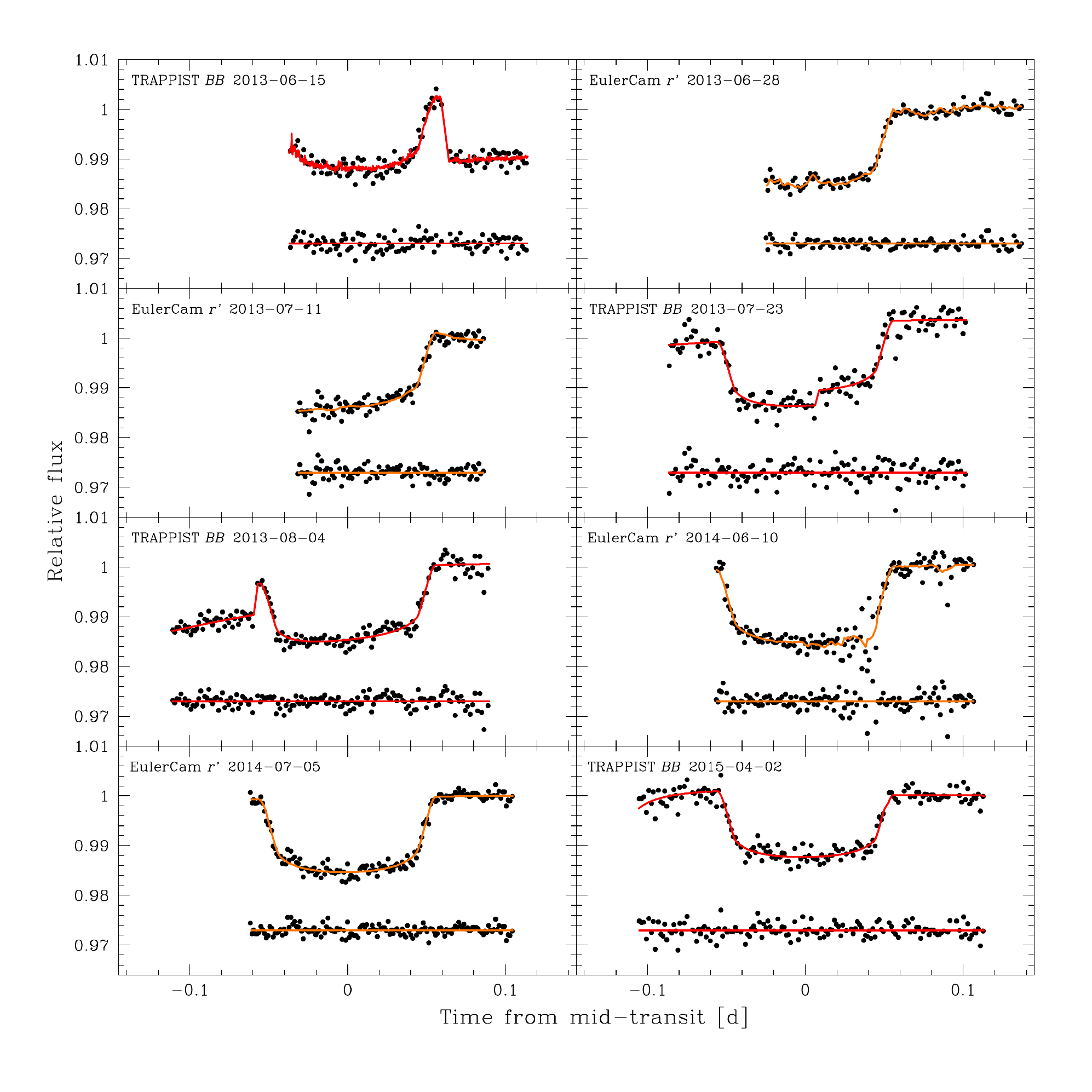} 
\vspace{-1.5cm}
\caption{All raw TRAPPIST and Euler/EulerCam transit light curves used in this work. The data are period-folded on the best-fit transit ephemeris from our global MCMC analysis (see Section \ref{analysis_method}) and binned in 2min bins. For each light curve, the overplotted solid line is the best-fit full model (photometric baseline $\times$ transit). The data and models are not corrected for the dilution by the nearby star here. The first five light curves were published in \citetalias{gillon14}, while the last three ones are new data.}     
\label{transits1}
\end{figure*}

\begin{figure*}
\centering     
\includegraphics[scale=0.90] {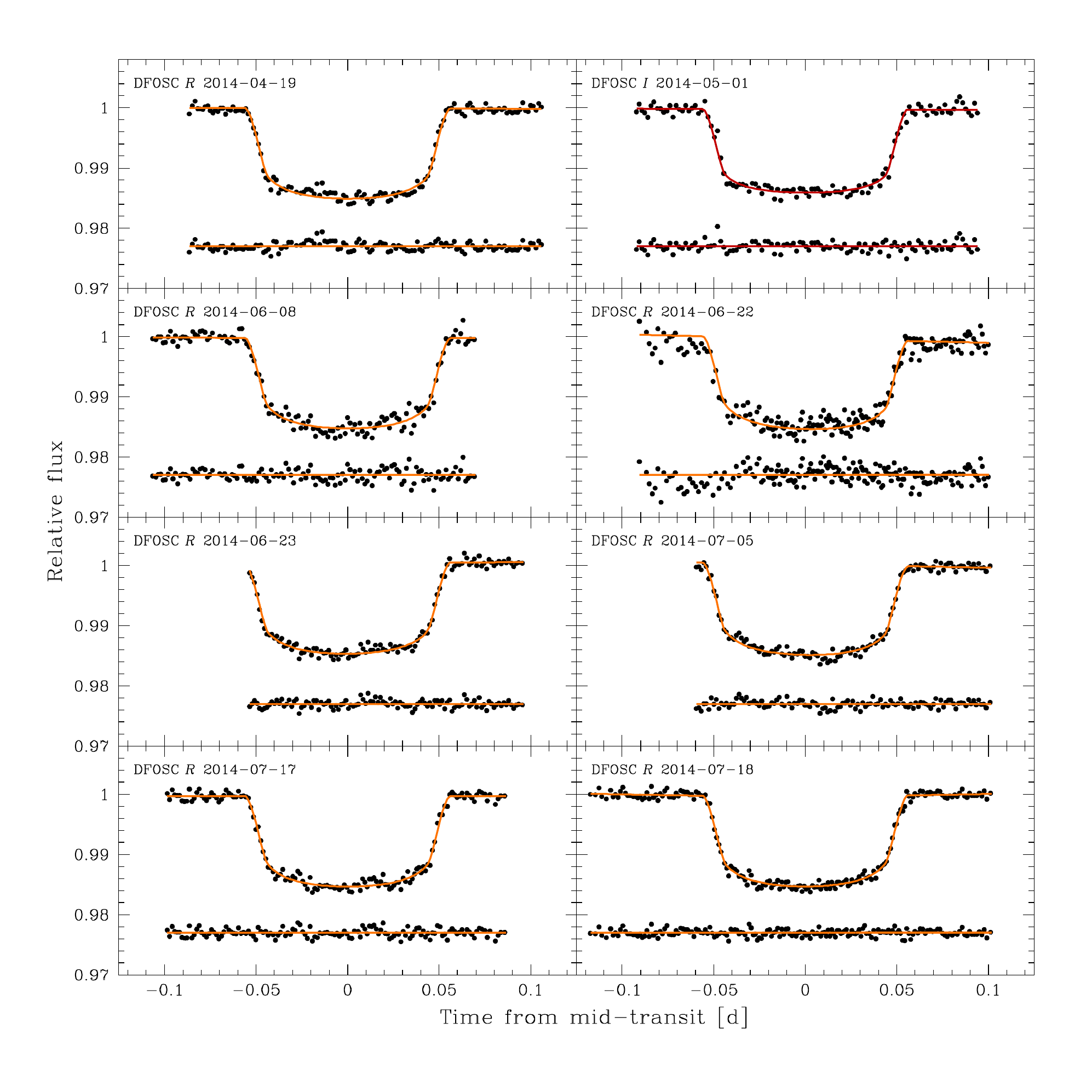} 
\vspace{-1.5cm}
\caption{All raw Danish/DFOSC transit light curves used in this work. The data are period-folded on the best-fit transit ephemeris from our global MCMC analysis (see Section \ref{analysis_method}) and binned in 2min bins. For each light curve, the overplotted solid line is the best-fit full model (photometric baseline $\times$ transit). The data and models are not corrected for the dilution by the nearby star here. All these light curves were published in \citetalias{southworth15}.}     
\label{transits2}
\end{figure*}

\begin{figure*}
\centering     
\includegraphics[scale=0.90] {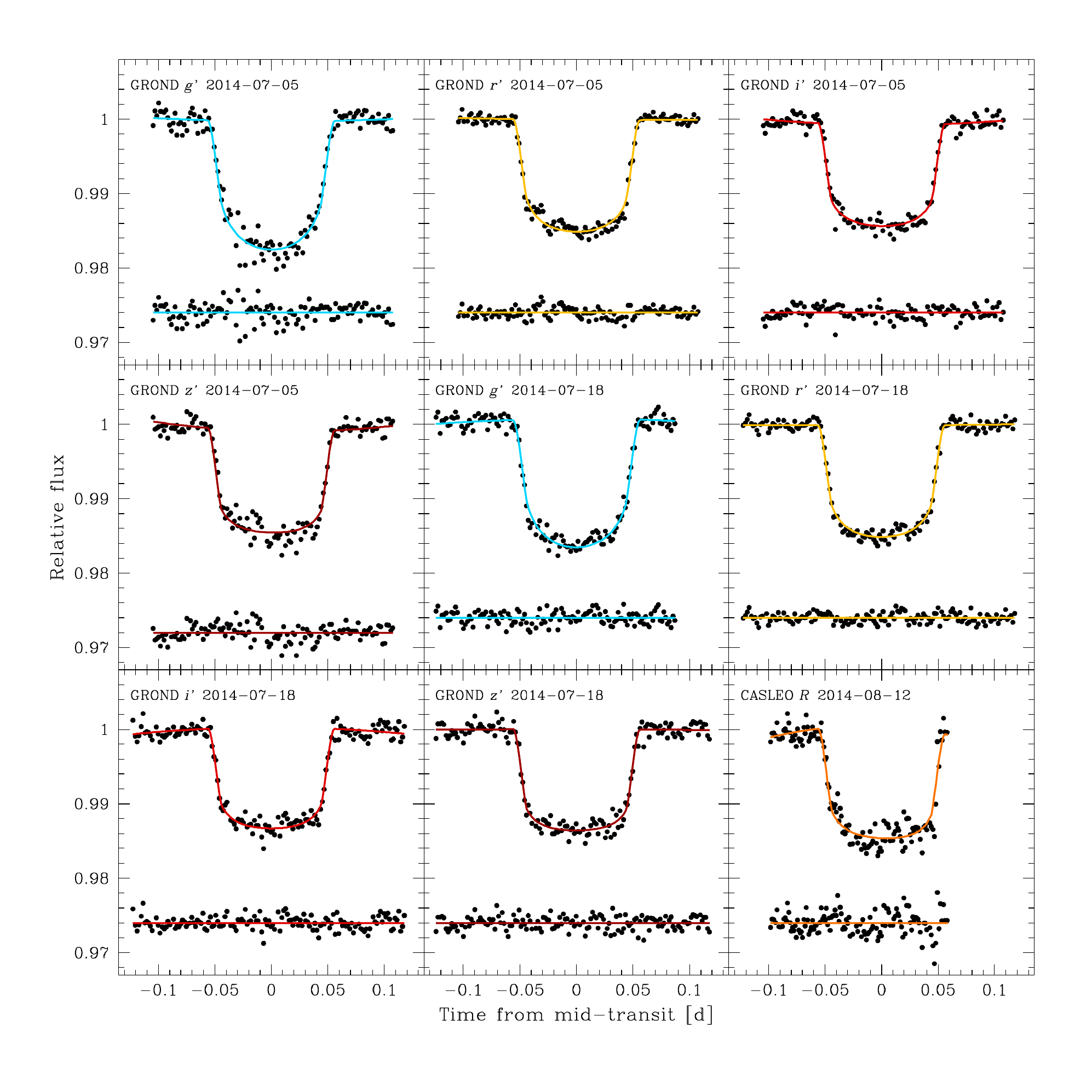} 
\vspace{-1.5cm}
\caption{All raw 2.2m/GROND and CASLEO/2.15m transit light curves used in this work. The data are period-folded on the best-fit transit ephemeris from our global MCMC analysis (see Section \ref{analysis_method}) and binned in 2min bins. For each light curve, the overplotted solid line is the best-fit full model (photometric baseline $\times$ transit). The data and models are not corrected for the dilution by the nearby star here. All these light curves were published in \citetalias{southworth15}.}     
\label{transits3}
\end{figure*}

%%%%%%%%%%%%%%%%%%%%%%%%%%%%%%%%%%%%%%%%%%%%%%%%%%

% Don't change these lines
\bsp	% typesetting comment
\label{lastpage}
\end{document}